\documentclass[twocolumn,usenatbib]{mn2e}

\usepackage{amsmath,amssymb,bm,float,dcolumn,Macro,mathrsfs,subfigure,tabularx}
\usepackage[dvips]{graphicx}
\usepackage{color}

\def\grad{\phi}
\def\curl{\varpi}

\def\N1{N$_1$-bias}

\def\ol{\overline}

\def\bl{\bm{\ell}}
\def\bL{\bm{L}}
\newcommand\Int[1]{\int\frac{d^2 #1}{(2\pi)^2}}
\def\bn{\bm{\nabla}}

\def\mcC{C}

\title{Bias-Hardened CMB Lensing}

\author[T. Namikawa et al.]
  {
  Toshiya Namikawa$^1$\thanks{E-mail: namikawa@utap.phys.s.u-tokyo.ac.jp}, 
  Duncan Hanson$^{2,3}$, 
  and Ryuichi Takahashi$^4$ \\ 
  $^1$Department of Physics, Graduate School of Science, The University of 
  Tokyo, Tokyo 113-0033, Japan \\ 
  $^2$Jet Propulsion Laboratory, California Institute of Technology,
  4800 Oak Grove Drive, Pasadena CA 91109, USA \\ 
  $^3$Department of Physics, McGill University, Montreal QC H3A 2T8, Canada\\
  $^4$Faculty of Science and Technology, Hirosaki University, 3 bunkyo-cho, 
  Hirosaki, Aomori, 036-8561, Japan
  }

\begin{document}

\maketitle
\begin{abstract}
We present  new methods for lensing reconstruction from CMB temperature fluctuations which 
have smaller mean-field and reconstruction noise bias corrections than current lensing 
estimators, with minimal loss of signal-to-noise.
These biases are usually corrected using Monte Carlo simulations, and to the extent that these 
simulations do not perfectly mimic the underlying sky there are uncertainties in the 
bias corrections.
The bias-hardened estimators which we present can have reduced sensitivity to such uncertainties, 
and provide a desirable cross-check on standard results.
To test our approach, we also show the results of lensing reconstruction from simulated temperature 
maps given on $10\times10$ deg$^2$, and confirm that our approach works well to reduce 
biases for a typical masked map in which $70$ square masks each having $10'$ on a side exist, 
covering $2$\% of the simulated map, which is similar to the masks used in the current SPT lensing 
analysis. 
\end{abstract} 



\section{Introduction}
\label{sec.1}

Ongoing, upcoming and next-generation CMB experiments are able to measure arcminute-scale 
temperature anisotropies, which are perturbed significantly by gravitational lensing.
Recently, several studies have reported the detection of lensing signals by reconstructing lensing 
fields involved in the CMB anisotropies, using the cross-correlations between CMB and large-scale 
structure \citep{Smith07,Hirata:2008cb,Bleem:2012gm,Sherwin:2012mr}, 
or CMB maps alone \citep{Das:2011ak,vanEngelen:2012va}. 
The measurement of  lensing effects with upcoming and next-generation CMB experiments
will be a powerful probe of the properties of dark energy and massive neutrinos 
(e.g., \citealt{Hu:2001fb,Lesgourgues:2006nd,dP09,Namikawa:2010re}), primordial non-Gaussianity 
(e.g., \citealt{Jeong:2009wi,Takeuchi:2011ej}), and cosmic strings 
(e.g.,\citealt{Yamauchi:2011cu,Namikawa:2011cs,Yamauchi:2012bc}). 

The distortion effect of lensing on the primary temperature anisotropies is expressed by a remapping. 
Denoting the primary temperature anisotropies at position $\hatn$ on the last scattering surface, 
$\Theta(\hatn)$, the lensed anisotropies in a direction $\hatn$, are given by 
\al{
	\tilde{\Theta}(\hatn) = \Theta(\hatn+\bm{d}(\hatn)) 
		= \Theta(\hatn) + \bm{d}(\hatn)\cdot\bm{\nabla}\Theta(\hatn) + \mcO(|\bm{d}|^2)
	\,. \label{remap}
}
The vector, $\bm{d}(\hatn)$, is the deflection angle, and, in terms of parity, we can decompose it 
into two terms, these are the gradient (even parity) and curl (odd parity) modes \citep{Namikawa:2011cs}: 
\al{
	\bm{d}(\hatn) = \bn \grad(\hatn) + (\star\bn)\curl(\hatn)
	\,, 
}
where the symbol, $\star$, denotes an operation which rotates the angle of a two-dimensional vector 
counterclockwise by 90 degrees. 
Hereafter, we refer to $\grad$ and $\curl$ as the scalar and pseudo-scalar lensing potential, respectively. 

Estimators for the lensing deflection field have been derived by several authors
(e.g., \citealt{Zaldarriaga:1998te,Seljak:1998aq,Hu:2001kj, Okamoto:2003zw,Hirata:2003ka,Bucher:2010iv,Namikawa:2011cs}).
These estimators all utilize the fact that a fixed lensing potential introduces statistical anisotropy 
into the observed CMB, in the form of a correlation between the CMB temperature and its gradient. 
With a large number of observed CMB modes, this correlation may be used to form estimates of the 
lensing potential.
The power spectrum of the lensing potential may in turn be studied by taking the power spectrum 
of these lensing estimates.

When performing lens reconstruction on a realistic dataset, there are two sources of bias which 
must be corrected for:
\begin{enumerate}
\item Statistical anisotropy from non-lensing sources such as a sky mask, inhomogeneous map noise, 
and asymmetry of the instrumental beam can be misinterpreted as lensing, generating a spurious lensing 
``mean-field''.
\item In the power spectrum of the lensing estimates, there can be a large contribution from the 
reconstruction noise which must be subtracted.
\end{enumerate}
With perfect statistical understanding of the CMB, foregrounds, and instrumental response to the sky 
these sources of bias are always computable (although potentially only through the use of Monte Carlo 
simulations), and may be subtracted from the lensing potential and power spectrum estimates. 
The bias terms are often quite large in comparison to the lensing signals of interest, however,
the result of which is that uncertainties in the power spectrum of the CMB fluctuations, the shape of 
the instrumental beam and its transfer function, the contribution from unresolved (and therefore unmasked) 
point sources, and the instrumental noise level can all lead to problematic uncertainties for these bias terms. 
Multiple approaches have been proposed in the literature to mitigate these problems.
For mean-field biases, most studies have focused on the issue of masking. 
One approach is to explicitly avoid masked regions \citep{Hirata:2008cb,Carvalho:2011gx}. 
Another is inpainting, in which masked regions are filled with simulated signal, therefore reducing the large 
gradients at the mask boundary \citep{Perotto:2009tv, Plaszczynski:2012ej}. 
Published analyses from the ACT and SPT experiments have used either source subtraction \citep{Das:2011ak} 
or inpainting via Wiener filtering \citep{vanEngelen:2012va} to deal with resolved point sources, and 
apodization to reduce spurious gradients at the survey boundary.
The reduction of reconstruction noise bias has also been discussed in the literature. 
For full-sky coverage, with homogeneous map noise and a symmetric beam the reconstruction noise 
bias may be estimated directly from the power spectrum of the map \citep{Hu:2001fa,Dvorkin:2008tf}. 
This approach has the added benefit of suppressing terms in the covariance of the reconstructed power 
spectrum \citep{Hanson:2010rp}.
For several specific choices of apodization/inpainting, it has been found that the full-sky equations can 
be quite accurate even for cut-sky data 
\citep{Das:2011ak,Plaszczynski:2012ej,vanEngelen:2012va,Feng:2011jx,Feng:2012uf}. 
In principle, the reconstruction noise bias may be avoided entirely by taking the cross-spectrum of two 
estimators with independent noise realizations \citep{Hu:2001fa}. 
This is a more difficult proposition than for power spectrum, where the only source of noise is 
instrumental and independent surveys of the same region of sky may usually be obtained. 
In the case of lensing, a large fraction of the reconstruction noise comes from the CMB fluctuations 
themselves, and so the construction of estimators with independent noise realizations requires slicing 
the Fourier plane into disjoint regions, such as the odd/even parity split \citep{Hu:2001fa} or the 
in/out split \citep{Sherwin:2010ge}. 
There is usually a substantial loss of signal-to-noise associated with such splits.
Additionally, for a realistic observation, some mode mixing is induced by the sky mask and instrumental 
response, and it is necessary to introduce buffer regions between the disjoint pieces of the Fourier plane 
from which the lensing estimates are formed, leading to further degradation of signal-to-noise.

In this paper, we discuss new methods for constructing lensing estimators which have significantly 
reduced mean-field and reconstruction noise biases, with minimal loss of signal-to-noise. 
These ``bias-hardened'' estimators can be constructed in conjunction with any of the data processing 
methods (such as inpainting, apodization, inverse-variance filtering) described above.
They are useful to deal not only with the complications induced by masking, but also
with other effects such as noise inhomogeneity, beam asymmetry, and uncertainty in
the primary CMB power spectrum.

This paper is organized as follows.
In Sec.~\ref{sec.2}, we briefly summarize quadratic estimators for lensing potentials from the 
CMB temperature, and the biases which they must be corrected for.
In Sec.~\ref{sec.3}, we propose methods for avoiding these biases.
In Sec.~\ref{sec.4}, we perform numerical simulations to test how well these methods work for 
several choices of map filtering and characteristic sky cuts. 
Section \ref{sec.5} is devoted to summary.

\section{Lensing reconstruction}
\label{sec.2} 

In this section, we briefly review the general formalism for quadratic temperature lensing estimators 
on the flat-sky \citep{Hu:2001kj,Cooray:2005hm,Namikawa:2011cs}.

\subsection{Lensing potential estimators} 

The statistical anisotropy introduced by lensing of the CMB is given by 
\al{
	\ave{\tilde{\Theta}_{\bl - \bL}\tilde{\Theta}_{\bL}}\rom{CMB}  
		= \delta_{\bl} \tilde{C}_L^{\Theta\Theta} + \sum_{x=\grad,\curl} f^x_{\bl,\bL} x_{\bl} 
	\,, 
}
Here, to distinguish from the ensemble average over realizations of the observed map including 
noise $\ave{\cdots}$, we denote $\ave{\cdots}\rom{CMB}$ as the ensemble average with a fixed 
scalar and pseudo-scalar mode of lensing potential. 
The weight function is given by
\al{
	f^x_{\bl,\bL} = \tilde{C}_L^{\Theta\Theta}\bl\odot_x\bL + \tilde{C}_{|\bl-\bL|}^{\Theta\Theta}\bl\odot_x(\bl-\bL) 
	\,, \label{sec2:weight}
}
where $x=\grad$ or $\curl$, and the operator, $\odot_x$, is defined, for arbitrary two vectors, 
$\bm{a}$ and $\bm{b}$, as 
\al{
	\bm{a}\odot_{\grad}\bm{b} = \bm{a}\cdot\bm{b} 
	\,, \qquad 
	\bm{a}\odot_{\curl}\bm{b} = (\star\bm{a})\cdot\bm{b} 
	\,. 
}
We note that the weight given in Eq.~\eqref{sec2:weight} uses the {\it lensed} power spectrum 
instead of {\it unlensed} power spectrum, following \cite{Lewis:2011fk}. 

This motivates the following generic form for a quadratic lensing estimator:
\al{
	\hat{x}^{\rm C}_{\bl} &= \hat{x}^{\rm S}_{\bl} - \langle\hat{x}^{\rm S}_{\bl}\rangle 
	\,, \label{est}
}
where the first term is the ``standard'' quadratic estimator 
\al{
	\hat{x}^{\rm S}_{\bl} & = \frac{1}{2}A_{\ell}^{x}\Int{\bL}f^x_{\bl,\bL}\bar{\Theta}_{\bL}\bar{\Theta}_{\bl-\bL} 
	\,, \label{est-x}
}
and a correction for the mean-field bias is given by the second term. 
Here $A_{\ell}^{x}$ is a normalization\footnote{In principle the normalization here should be 
a matrix, however for most realistic situations this is impractical and instead an \textit{effective} 
normalization like the one above is used.}, and $\bar{\Theta}_{\bL}$ are the Fourier modes of a 
filtered sky map. 

The filtered sky map $\bar{\Theta}_{\bL}$ may be obtained in a variety of different ways. 
Inverse-variance (or ``$C^{-1}$'') filtering can be shown to minimize the reconstruction noise 
\citep{Hirata:2002jy,Smith07,Hanson:2009gu}. 
For a sky map $\hat{\Theta}(\hatn)$ with homogeneous noise and a delta function instrumental beam, such that the data model is given by
\al{
	\hat{\Theta}(\hatn) = \Theta(\hatn) + n(\hatn) 
	\,, \label{eqn:datamodel}
}
where $\Theta$ is a (lensed or unlensed) CMB realization and $n(\hatn)$ is a noise realization, 
the corresponding diagonal filter is given by
\al{
	\bar{\Theta}_{\bL} = \frac{1}{\hat{\mcC}_L^{\Theta\Theta}} \hat{\Theta}_{\bL} 
	\qquad {\text{ (diagonal)}} 
	\,. \label{eqn:diagfilt}
}
The quantity $\hat{\mcC}_L^{\Theta\Theta}$, is the theoretical ensemble average angular 
power spectrum of the observed sky, including the contribution from instrumental noise. 
This diagonal filter is simple to implement, and can be used even on the masked sky, with the 
penalty of large spurious gradients at the mask boundary \citep{Hirata:2008cb}. 
Intermediate between full $C^{-1}$ filtering and the diagonal approximation is the approach of 
using an apodized sky mask to reduce the creation of spurious gradients at the mask boundary. 
In this paper we will study all three approaches, in conjunction with bias-hardened estimators.
In all of our discussion, we will assume that whichever filter is chosen, in regions far from any
mask boundary it asymptotes to the form given in Eq.~\eqref{eqn:diagfilt}.

On the full sky, with diagonal filtering, the estimator normalization may be determined analytically 
and is given by
\al{
	A_{\ell}^{x} &= \left\{\Int{\bL}\frac{(f^x_{\bl,\bL})^2}{2\hat{\mcC}_L^{\Theta\Theta}\hat{\mcC}_{|\bl-\bL|}^{\Theta\Theta}}\right\}^{-1} 
	\,. \label{GN}
}
In more general situations this expression does not necessarily hold and the normalization must 
always be determined using Monte Carlo.
It is still incredibly useful, however, if the effective normalization is close to that given by 
Eq.~\eqref{GN}, as this equation may be used to propagate uncertainties on the weight function of 
Eq.~\eqref{sec2:weight} (for example, due to uncertainties in the instrumental beam transfer function) 
to uncertainties on the normalization without the need for expensive simulations.

A numerical approach to fast computation of the generic quadratic estimator is to rewrite Eq.~\eqref{est-x} 
as a convolution of two maps, $\bar{\Theta}_{\bl}$, and $\bm{\alpha}_{\bl}=i\bl\,\tilde{C}_{\ell}^{\Theta\Theta}\bar{\Theta}_{\bl}$ 
\citep{Hu:2001kj}: 
\al{
	\hat{x}^{\rm S}_{\bl} &= A_{\ell}^{x}\Int{\bL}\bar{\Theta}_{\bL}[\bl\odot_x\bm{\alpha}_{\bl-\bL}]  
	\,. \label{eqn:realspace}
}
With the convolution theorem, in real space this can be written as
\al{ 
	\hat{x}^{\rm S}_{\bl} &= A_{\ell}^{x} \int d^2\hatn \,e^{-i\bl\cdot\hatn}\bar{\Theta}(\hatn)[\bl\odot_x\bm{\alpha}(\hatn)]  
	\,, \label{conv}
} 
where the quantities, $\bar{\Theta}(\hatn)$ and $\bm{\alpha}(\hatn)$, are the real-space counterpart 
of $\bar{\Theta}_{\bl}$ and $\bm{\alpha}_{\bl}$, respectively. 
Eq.~\eqref{conv} means that the estimator can be computed by Fast Fourier Transform.

\subsection{Power spectrum estimators} 
\label{subsec:powerspectrumestimators}

The power spectrum of the lensing potential may be studied through the power spectra of the 
quadratic estimators above.
This quantity probes the 4-point function of the lensed CMB, and can be usefully broken into 
disconnected and connected parts as
\al{
	\ave{|\hat{x}^{\rm C}_{\bl}|^2} 
		= \ave{|\hat{x}^{\rm C}_{\bl}|^2}_D +  \ave{|\hat{x}^{\rm C}_{\bl}|^2}_C
	\,. \label{pow}
}
The disconnected part, $\ave{\cdots}_D$, contains the contributions which would be expected if 
$\bar{\Theta}_{\bL}$ were a Gaussian random variable, while the connected part, $\ave{\cdots}_C$, 
contains the non-Gaussian contributions which are a distinctive signature of lensing.
The disconnected part represents the reconstruction noise bias discussed earlier, which must be 
accurately subtracted from Eq.~\eqref{pow} to obtain a clean measurement of the lensing signal. 

The disconnected bias is given by
\begin{multline}
	\ave{|\hat{x}^{\rm C}_{\bl}|^2}_D = \frac{1}{2}\left( A_{\ell}^{x} \right)^2\Int{\bL}\Int{\bL'}
	\\ \times 
	f^x_{\bl,\bL} f^x_{\bl,\bL'}\bar{C}_{ \bL, \bl-\bL' } \bar{C}_{\bl-\bL, \bL'},
\label{eqn:disconnectedbias}
\end{multline}
where $\bar{C}_{\bL, \bL'}\!=\!\ave{\bar{\Theta}_{\bL}\bar{\Theta}_{\bL'}}$ is the covariance 
matrix of the filtered map. 
Given a model for this covariance matrix and a method to simulate Gaussian realizations of it, 
this disconnected bias may be evaluated by Monte Carlo.
If the covariance matrix is diagonal then it is tractable to evaluate the disconnected bias analytically. 
For the filtering of Eq.~\eqref{eqn:diagfilt}, the disconnected bias is equal to the normalization, 
with $A_{\ell}^{x} = \ave{|\hat{x}^{\rm C}_{\bl}|^2}_D \equiv N_{\ell}^{x,(0)}$, where $A_{\ell}^{x}$ 
is given by Eq.~\eqref{GN}.

The power spectrum of the connected part of the quadratic estimator is given on the full sky by 
\citep{Kesden:2003cc}
\al{
	\ave{|\hat{x}^{\rm C}_{\bl}|^2}_C = C_{\ell}^{xx} + N_{\ell}^{x,(1)} 
	\,. \label{pow-c}
}
Here $C_{\ell}^{xx}$ is the potential power spectrum which it was our intention to reconstruct, 
while $N_{\ell}^{x,(1)}$ is a nuisance term coming from the ``secondary'' lensing contractions of 
the trispectrum \citep{Hu:2001fa}.

\section{Bias-hardened estimators} 
\label{sec.3} 

In this section, we propose methods to mitigate the mean-field and reconstruction noise biases 
discussed in the previous section. 

\subsection{Bias-reduced lensing estimator} 
\label{sec:mres}

We begin by discussing a method to reduce the lensing mean-field.
Our approach is straightforward; we construct a new estimator which is optimized to detect the 
source of the mean-field bias, and use this estimate to correct the lensing estimator accordingly. 
This allows the construction of a new hybrid lensing estimator which has intrinsically much 
smaller mean-fields than the standard one. 

To illustrate this approach, it is useful to consider the mean-field bias introduced by a specific 
form of statistical anisotropy.
To start, consider a modulation of the observed temperature fluctuations by a function $W(\hatn)$:
\beq 
	\hat{\Theta}^{\rm mod}(\hatn) = W(\hatn) \hat{\Theta}(\hatn) 
	\,, \label{mask}
\eeq 
From Eq.~(\ref{mask}), the multipole coefficients of observed temperature anisotropies are 
related to the underlying fluctuations as 
\al{ 
	\hat{\Theta}^{\rm mod}_{\bl} 
		= \int d^2\hatn \, e^{i\bl\cdot\hatn} \, \hat{\Theta}^{\rm mod}(\hatn) 
		= \Int{\bl\p} W_{\bl-\bl\p} \hat{\Theta}_{\bl\p} 
	\,, \label{fourier-W}
} 
where we use the Fourier counterpart of the window function: 
\al{
	W_{\bl} &= \int d^2\hatn \, e^{i\bl\cdot\hatn} \, W(\hatn) 
	\,. 
}
Introducing a mask function $M_{\bl}=\delta_{\bl}-W_{\bl}$, we can rewrite this as
\al{ 
	\hat{\Theta}^{\rm mod}_{\bl} = \hat{\Theta}_{\bl} - \Int{\bl\p} M_{\bl-\bl\p} \hat{\Theta}_{\bl\p} 
	\,. \label{fourier-M}
} 
The covariance matrix of the masked sky is then given by
\al{
	\langle \hat{\Theta}^{\rm mod}_{\bL}\hat{\Theta}^{\rm mod}_{\bl-\bL}\rangle \rom{CMB,n} 
		&= \delta_{\bl}\hat{\mcC}_L^{\Theta\Theta} + \sum_{x=\grad,\curl} f^x_{\bl,\bL} x_{\bl} 
		\notag \\ 
		&\qquad + {f}_{\bl,\bL}^M M_{\bl} + {\cal O}(M^2)
	\,, \label{off-diag}
}
where we define ${f}^M_{\bl,\bL}=-\hat{\mcC}^{\Theta\Theta}_{L}-\hat{\mcC}^{\Theta\Theta}_{|\bl-\bL|}$. 
Note here that $\ave{\cdots}\rom{CMB,n}$ means the ensemble average over the realizations of 
CMB and noise with a fixed realizations of lensing potentials. 
At first order in $M$, with diagonal filtering, Eq.~(\ref{off-diag}) leads to the following bias for 
the scalar lensing potential estimator:
\al{
	\ave{\hat{\grad}_{\bl}^{\rm S}}\rom{CMB,n} = \grad_{\bl} + A_{\ell}^{\grad}\left[
		\Int{\bL}\frac{{f}^{\grad}_{\bl,\bL}{f}^M_{\bl,\bL}}
		{2\hat{\mcC}^{\Theta\Theta}_{L}\hat{\mcC}^{\Theta\Theta}_{|\bl-\bL|}} \right] M_{\bl}
	\,. \label{maskbias}
}
We see that masking introduces a mean-field which directly traces the mask $M_{\bl}$. 
Unlike the scalar lensing potential, the estimator for pseudo-scalar lensing potential is unmodified 
at first order in $M$. 
Masking does not introduce a large mean-field into the pseudo-scalar lensing potential.

An estimator for the mask field, $M_{\bl}$, can be constructed analogous to that for lensing:
\al{
	\hat{M}^{\rm S}_{\bl} = \frac{1}{2}A_{\ell}^{M} \Int{\bL} f_{\bl,\bL}^{M}\bar{\Theta}_{\bL}\bar{\Theta}_{\bl-\bL} 
	\,, \label{maskest-1}
}
where $A_{\ell}^{M}$ is the same as Eq.~(\ref{GN}) but using the mask weight function, 
${f}_{\bl,\bL}^M$. 

Now we consider the joint estimation of both the mask and lensing fields simultaneously. 
The standard quadratic estimator is biased by masking as Eq.~(\ref{maskbias}). 
Correspondingly, the naive mask estimator of Eq.~(\ref{maskest-1}) is biased by lensing as 
\al{
	\ave{\hat{M}_{\bl}^{\rm S}}\rom{CMB,n} = M_{\bl} + \left[{A}_{\ell}^{M}\Int{\bL} 
		\frac{{f}_{\bl,\bL}^{M}{f}_{\bl,\bL}^{\grad}}
		{2\hat{\mcC}_L^{\Theta\Theta}\hat{\mcC}_{|\bl-\bL|}^{\Theta\Theta}}\right]\grad_{\bl}
	\,. \label{maskest-2}
}
In matrix form, we can write (temporarily ignoring the mean-field corrections for both estimators)
\al{
	\begin{pmatrix} \ave{\hat{\grad}^{\rm S}_{\bl}}\rom{CMB,n} \\ 
		\ave{\hat{M}_{\bl}^{\rm S}}\rom{CMB,n} \end{pmatrix} 
		= \begin{pmatrix} 1 & R^{\grad,M}_{\ell} \\ R_{\ell}^{M,\grad} & 1 \end{pmatrix} 
		\begin{pmatrix} \grad_{\bl} \\ M_{\bl} \end{pmatrix} 
	\,, \label{mat}
}
where the ensemble average is taken over CMB and noise realizations and we define the response 
function $R_{\ell}^{a,b}$, for $a,b=\grad,M$, 
\al{
	R_{\ell}^{a,b} = A_{\ell}^{a}\Int{\bL} \frac{{f}_{\bl,\bL}^{a}{f}_{\bl,\bL}^{b}}
		{2\hat{\mcC}_L^{\Theta\Theta}\hat{\mcC}_{|\bl-\bL|}^{\Theta\Theta}}
	\,. \label{eqn:responsefunc}
}
By inverting Eq.~(\ref{mat}), we obtain 
\al{
	\begin{pmatrix} \grad_{\bl} \\ M_{\bl} \end{pmatrix} 
		= \frac{1}{1-R^{x,M}_{\ell}R^{M,x}_{\ell}}
		\begin{pmatrix} 1 & - R^{x,M}_{\ell} \\ - R_{\ell}^{M,x} & 1 \end{pmatrix} 
		\begin{pmatrix} \ave{\hat{\grad}^{\rm S}_{\bl}}\rom{CMB,n} \\ 
			\ave{\hat{M}^{\rm S}_{\bl}}\rom{CMB,n} \end{pmatrix}
	\,. \label{sec3:mat}
}
A bias-free estimator for the scalar lensing potential may therefore be formed as
\al{
	\hat{\grad}^{\rm BR}_{\bl} = \frac{\hat{\grad}^{\rm S}_{\bl}- R_{\ell}^{x,M}\hat{M}^{\rm S}_{\bl}}
		{1-R^{x,M}_{\ell}R^{M,x}_{\ell}}
	\,. \label{BR-est}
}
This lensing estimator has no mean-field contribution from the mask (up to the $M^2$ term in Eq.~\ref{off-diag}).
Of course, in a practical situation the estimator of Eq.~\eqref{BR-est} will not be completely free of mask bias. 
The $M^2$ term may produce a mean-field contribution, and also if non-diagonal filtering of the 
map is utilized then the response terms of Eq.~\eqref{eqn:responsefunc} are only approximate. 
Even in such situations, however, the estimator of Eq.~\eqref{BR-est} should have a smaller mask 
mean-field than the standard estimator. 
We therefore refer to this as a ``bias-reduced (lensing) estimator''. 
In Sec.~\ref{sec.4} we will test the behavior of this estimator for several choices of filtering and mask.
The procedure above can be easily generalized to mitigate multiple sources of mean-field simultaneously. 

In the above, although we focus on masking, the biases from inhomogeneous map noise and beam asymmetry are similar.
For inhomogeneous noise, the instrumental noise in Eq.~(\ref{eqn:datamodel}) has non-zero 
off-diagonal terms in its covariance, and leads to the mean-field bias given by \citep{Hanson:2009dr}
\al{
	\ave{\hat{\grad}_{\bl}^{\rm S}} 
		= A_{\ell}^{\grad}\left[
		\Int{\bL}\frac{{f}^{\grad}_{\bl,\bL}{f}^N_{\bl,\bL}}
		{2\hat{C}^{\Theta\Theta}_{L}\hat{C}^{\Theta\Theta}_{|\bl-\bL|}} \right] N_{\bl}
	\,, \label{inhomobias}
}
where $N_{\bl}$ is the Fourier transform of $n^2(\hatn)$, and the weight function is $f_{\bl,\bL}^N=1$. 
With Eq.~(\ref{inhomobias}), the bias-reduced estimator for inhomogeneous noise can be constructed analogous 
to that for masking. 
For beam asymmetry, with the beam response, $r(\hatn,\hatn\p)$, we rewrite Eq.~(\ref{eqn:datamodel}) in more general form 
(e.g. \citealt{Souradeep:2001ds,Hanson:2010gu}): 
\al{
	\hat{\Theta}^{\rm mod}(\hatn) = \int d^2\hatn\p r(\hatn,\hatn\p) \Theta(\hatn\p) + n(\hatn)
	\,. \label{beam-real}
}
The beam transfer function, $B_{\ell,s}$, is defined as 
\al{
	r(\hatn,\hatn\p) = \sum_{s} \Int{\bl} B_{\ell,s} e^{i(\hatn-\hatn\p)\cdot\bl} 
		e^{is(\varphi_{\ell}-\varphi-\psi(\hatn))} 
	\,. 
}
With $b_{\ell,s}=B_{\ell,s}/B_{\ell,0}$, Eq.~(\ref{beam-real}) is then rewritten as \citep{Hanson:2010gu} 
\beq 
	\hat{\Theta}^{\rm mod}(\hatn) = \hat{\Theta}(\hatn) + 
		\sum_{s\not=0} \int \frac{d^2\bl}{(2\pi)^2} b_{\ell,s} \Theta_{\bl} e^{i\bl\cdot\hatn} 
		e^{is(\varphi_{\ell}-\varphi-\psi(\hatn))} 
	\,, \label{beam} 
\eeq 
where $\varphi_{\ell}$ and $\varphi$ are azimuthal angle of $\bl$ and $\hatn$, 
$\psi(\hatn)$ denotes the angle of the fiducial beam axis, and the symmetric part is assumed to be $B_{\ell,0}=1$. 
Assuming that the asymmetric terms of normalized beam transfer function, $b_{\ell,s} (s\not=0)$, are small, this leads 
to the following mean-field bias \citep{Hanson:2010gu}: 
\al{
	\ave{\hat{\grad}_{\bl}^{\rm S}} 
		&= \sum_{s\not=0} A_{\ell}^{\grad}\left[\Int{\bL}\frac{{f}^{\grad}_{\bl,\bL}{f}^{B}_{s,\bl,\bL}} 
		{2\hat{C}^{\Theta\Theta}_{L}\hat{C}^{\Theta\Theta}_{|\bl-\bL|}} \right] \psi_{\bl,s} 
	\,, \label{beambias}
}
where $\psi_{\bl,s}$ is the spin-weighted Fourier counterpart of $e^{-is\psi(\hatn)}$ and we define 
\al{
	{f}^{B}_{s,\bl,\bL} = b_{L,s}\hat{C}_{L}^{\Theta\Theta}	+ (\bL \leftrightarrow \bl-\bL) 
	\,. 
}

Note here that, 
the effects of symmetric part of beam transfer function, 
which is ignored in the discussions of mean-field bias above, 
modify the weight functions described in the above. 
This can be done simply by taking $f_{\bl,\bL} \rightarrow f_{\bl,\bL}B_{L,0}^{-1}B_{|\bl-\bL|,0}^{-1}$. 

In the case of masking (as well as beam asymmetries and noise inhomogeneity), the shape of the 
mask
-field $M_{\bl}$ is already known perfectly and we would also subtract the 
mean-field bias using the naive subtraction of Eq.~(\ref{est}). 
However, the bias-reduced estimator still has importance since it is the optimal unbiased estimator 
in the presence of mask field, i.e., the above estimator is uniquely determined by imposing 
the unbiased and optimal conditions on both the lensing and mask estimators, in the presence of 
both lensing and masking effects 
\footnote{This is the same analogy of \citet{Namikawa:2011cs}, but now we consider the 
lensing and mask fields, which do not separately estimate each other. 
Thus, we have to consider the response induced by the other, as shown in Eq.~(\ref{eqn:responsefunc}).
}. 

Our intention for the construction of bias-reduced estimators is to mitigate the effect of possible 
errors in our understanding of the mean-field. 
Even if we know the mean-field $M_{\bl}$ perfectly, there are other sources of possible error 
for the mean-field subtraction of Eq.(\ref{est}). 
Suppose, for example, that our analysis is performed using a slightly incorrect estimate 
$\hat{\cal C}_L^{\Theta\Theta}$ (denoted by a cursive $\cal C$) for the ensemble average map 
power spectrum $\hat{C}_L^{\Theta\Theta}$, with 
\al{
	\hat{\cal C}_L^{\Theta\Theta} = \hat{\mcC}_L^{\Theta\Theta} - \Sigma_L
	\,. \label{eq:powspecerr}
}
This propagates directly to an error $f^{\Sigma}_{\bl, \bL}=-\Sigma_L-\Sigma_{|\bl-\bL|}$ in the 
weight function of Eq.~\eqref{off-diag}. 
If the mean-field bias is removed by averaging over masked CMB realizations with power spectrum 
given by Eq.~\eqref{eq:powspecerr}, there will be an uncorrected mean-field contribution given by 
\al{
	\ave{\grad^{\rm C}_{\bl}} = \mcR^{\grad, \Sigma} M_{\bl} 
	\,,	\label{eqn:sigmastandardbias}
}
where $\mcR^{ab}_{\ell}$ is the same as Eq.~(\ref{eqn:responsefunc}) but using the incorrect 
power spectrum. 
For the bias-reduced estimator, however, the uncorrected contribution is given by 
\al{
	\ave{\grad^{\rm BR}_{\bl}} = \frac{\mcR^{\grad, \Sigma} - \mcR^{\grad, M} \mcR^{M, \Sigma}}
		{1-\mcR^{\grad,M}_{\ell}\mcR^{M,\grad}_{\ell}} M_{\bl} 
	\,. \label{eqn:sigmamrebias}
}
It is in principle possible that the residual mean-field in this case is \textit{worse} than for the 
standard approach, for example if $\mcR^{\grad,\Sigma}$ is zero but $\mcR^{M,\Sigma}$ is not. 
If bounds on $\Sigma$ can be obtained for a specific experiment, the usefulness of the bias-reduced 
estimators can be explored using Eq.~\eqref{eqn:sigmastandardbias} and Eq.~\eqref{eqn:sigmamrebias}.
As an example, consider the case of a calibration error, 
$f^{\Sigma}_{\bl,\bL} = b f^{M}_{\bl,\bL}$, for some small coefficient $b$. 
In this case the residual mean-field will be completely avoided by the bias-reduced estimator. 
In any case, agreement between standard and bias-reduced estimators provides a useful consistency test.

\subsection{Noise bias estimator}

We turn now to the issue of reconstruction noise bias, given by Eq.~\eqref{eqn:disconnectedbias}, 
which is dependent on the covariance matrix of the filtered CMB modes $\bar{C}_{\bL, \bL'}$.
Similar to the case of the bias-reduced lensing estimator above, we suppose that we are in possession 
of an imperfect model $\bar{\cal C}_{\bL, \bL'}$ for the ensemble-average covariance matrix of 
the filtered CMB modes
\al{
	\bar{\cal C}_{\bL, \bL'}  = \bar{C}_{\bL, \bL'} - \Sigma_{\bL, \bL'} 
	\,, \label{eqn:imperfectmodel}
}
where $\Sigma_{\bL, \bL'}$ is an error matrix.
An estimate of the reconstruction noise made by substituting ${\cal C}$ for $C$ in 
Eq.~\eqref{eqn:disconnectedbias} will have ${\cal O}( \Sigma )$ contributions from the error matrix. 
We would therefore like to construct an estimator to determine the reconstruction noise bias more 
directly from the data. 

For full-sky coverage and diagonal filtering, where the covariance matrix is given by 
$\bar{C}_{\bL, \bL'} = \delta_{\bL-\bL'} \bar{C}_{\bL}$, this can be done simply, by replacing 
the ensemble average $\bar{C}_{\bL}$ in Eq.~\eqref{eqn:disconnectedbias} with the 
(realization-dependent) power spectrum of the filtered map \citep{Hu:2001fa,Dvorkin:2008tf}. 
This method of correcting the disconnected bias has the added advantage that it removes the largest 
off-diagonal contributions to the covariance matrix of the power spectrum estimates \citep{Hanson:2010rp}.
In more realistic situations where the covariance matrix has off-diagonal elements this procedure 
is not guaranteed to work, although for some specific forms of filtering it has been found adequate 
\citep{Das:2011ak,vanEngelen:2012va, Plaszczynski:2012ej}. 

Here we motivate a new approach which utilizes both data and the imperfect covariance. 
It is more robust than relying entirely on $\bar{\cal C}_{\bL, \bL'}$, and also does not depend on the accuracy of 
full-sky equations which neglect any off-diagonal correlations due to masking, inhomogeneity 
of the instrumental noise, etc.
The method is again straightforward, we simply estimate the reconstruction noise bias using 
Eq.~\eqref{eqn:disconnectedbias}, substituting the imperfect model for one of the covariance 
matrices and the data itself for the other:
\begin{multline}
	\widehat{ \ave{|\hat{x}_{\bl}|^2}}_D = \left( A_{\ell}^{x} \right)^2\frac{1}{2}\Int{\bL}\Int{\bL'}f^x_{\bl,\bL} f^x_{\bl,\bL'}
    \\ \times
	\left( 2 \bar{\cal C}_{ \bL, \bl-\bL' } \bar{\Theta}_{\bl-\bL} \bar{\Theta}_{\bL'}^{*} 
		- \bar{\cal C}_{ \bL, \bl-\bL' } \bar{\cal C}_{\bl-\bL, \bL'} \right)
	\,. \label{eqn:disconnectedbiashalfsim}
\end{multline}

This approach to removal of the disconnected bias emerges naturally when deriving optimal 
trispectrum estimators from an Edgeworth expansion of the CMB likelihood. 
In the Edgeworth expansion, the likelihood function of the CMB fluctuations at trispectrum order is given 
by (e.g. Eq.~(B3) of \cite{Regan:2010cn})
\al{
	\mcL/\mcL_G &= 1+3\left[\prod_{i=1}^4\Int{\bl_i}\right]T_{\bl_1,\bl_2,\bl_3,\bl_4} 
	\notag \\ 
	& \qquad \times \Big(\frac{\bar{\Theta}_{\bl_1}\bar{\Theta}_{\bl_2}\bar{\Theta}_{\bl_3}\bar{\Theta}_{\bl_4}}{3} 
	\notag \\ 
	& \qquad \qquad - 2\bar{C}_{\bl_1,\bl_2}\bar{\Theta}_{\bl_3}\bar{\Theta}_{\bl_4} 
		+ \bar{C}_{\bl_1,\bl_2}\bar{C}_{\bl_3,\bl_4}\Big) 
	\,, 
}
where $\mcL_G$ is the Gaussian likelihood, and 
$T_{\bl_1,\bl_2,\bl_3,\bl_4}=\ave{\tilde{\Theta}_{\bl_1}\tilde{\Theta}_{\bl_2}\tilde{\Theta}_{\bl_3}\tilde{\Theta}_{\bl_4}}\rom{c}$ is 
the trispectrum, which in this case is generated by lensing.
The maximum-likelihood estimator for the lensing power spectrum is obtained by 
maximizing the above likelihood, setting is derivative with respect to $C_{\ell}^{xx}$ to zero. 
Ignoring the lensing power spectrum in the weight function and covariance matrix, 
the differential operator with respect to $C_{\ell}^{xx}$ acts only on the trispectrum which is written as 
\al{ 
	\frac{\delta T_{\bl_1,\bl_2,\bl_3,\bl_4}}{\delta C_{\ell}^{xx}}
	&= f^x_{\bl,\bl_1}[f^x_{\bl,-\bl_3}\delta_{\bl-\bl_{12}}\delta_{\bl+\bl_{34}} 
	\notag \\ 
	&+ f^x_{\bl,\bl+\bl_2}(\delta_{\bl-\bl_{13}}\delta_{\bl+\bl_{24}} 
	+\delta_{\bl-\bl_{14}}\delta_{\bl+\bl_{23}})] 
	\,, 
} 
where $\bl_{ij}=\bl_i+\bl_j$. This leads to 
\al{
	\frac{\delta \ln\mcL}{\delta C_{\ell}^{xx}} 
	&= \frac{3\mcL_G}{\mcL}\Int{\bL}\Int{\bL\p} f^x_{\bl,\bL}f^x_{\bl,\bL\p} \notag \\ 
	&\times \bigg[\bar{\Theta}_{\bL}\bar{\Theta}_{\bl-\bL}(\bar{\Theta}_{\bL\p}\bar{\Theta}_{\bl-\bL\p})^* 
	\notag \\ 
	&- 2(2\bar{C}_{\bL,\bl-\bL\p}\bar{\Theta}_{\bl-\bL}\bar{\Theta}_{\bL\p}^* 
		- \bar{C}_{\bL,\bl-\bL\p}\bar{C}_{\bl-\bL,\bL\p})
	\bigg] \,, 
}
where we assume that the mean-field bias which is generated by $\bar{C}_{\bL,\bl-\bL}$ is completely subtracted. 
We find that the reconstruction noise estimator which comes from the last line of the above equation is 
then given by Eq.~(\ref{eqn:disconnectedbiashalfsim}). 

The calculation of the bias in this manner is only sensitive to uncertainties in the CMB covariance at 
${\cal O}( \Sigma^2 )$, an improvement over an entirely model-based determination of the 
reconstruction noise. 
It also maintains the property of suppressing off-diagonal contributions to the covariance matrix of 
the reconstructed power spectrum. 
We refer to the approach above as the ``noise bias estimator''.

\section{Numerical Tests} \label{sec.4} 

In this section, we test the usefulness of the bias-reduced estimator for several different choices of 
filtering and characteristic masking.
As discussed in Sec.~\ref{sec:mres}, we expect significant reduction of the mask mean-field bias using 
these estimators, which reduces the ability for some uncertainties in the primary CMB and instrument 
properties to leave residual biases in the estimated lensing potential. 
We do not test the noise bias estimator, as the only potential limitation of its usefulness is in the size of 
the $\Sigma^2$ matrix in Eq.~\eqref{eqn:imperfectmodel}, and this must be evaluated on an 
experiment-specific basis. 
We will, however, compare the noise bias and normalization for the bias-reduced estimators to the 
approximation of Eq.~\eqref{GN} as a cross-check on how closely they agree with the full-sky 
expectation for various choices of filtering, with or without the use of bias-reduced reduced estimators.

For lensing reconstruction, we use $100$ realizations of Gaussian unlensed and lensed temperature 
fluctuations. 
These fluctuations are simulated on a $10\times10$ deg$^2$ map, and the lensed maps are generated 
using the ray-tracing simulations described in Appendix~\ref{appA}.
We generate simulated source masks by cutting $N\rom{m}$ randomly located square regions with 
angular size $r\rom{m}$ on each side.
In our analysis, we choose $N\rom{m}=70$ and $r\rom{m}=10'$ or $20'$. 
Note that the case with $N\rom{m}=70$ and $r\rom{m}=10'$ roughly corresponds to the case of 
SPT lensing analysis \citep{vanEngelen:2012va}. 
To consider experiments with high-angular resolution such as PolarBear, ACTPol and SPTPol, and 
also to avoid contamination by SZ and unresolved point sources, we assume a delta function instrumental 
beam, but truncate the temperature multipoles at $\ell\rom{max}=3000$.
We assume homogeneous map noise, with a level of 1 $\mu$K arcmin. 
The angular power spectrum of the lensed/unlensed temperature and scalar lensing potential is computed 
using CAMB \citep{Lewis:1999bs}. 
Throughout this section, our fiducial values of cosmological parameters are consistent with WMAP 7-year 
results \citep{Komatsu:2010fb}; 
$\Omega\rom{b}=0.046$, $\Omega\rom{m}=0.27$, $\Omega_{\Lambda}=0.73$, $h=0.70$, 
$\sigma_8=0.81$, $n\rom{s}=0.97$.

\subsection{Filtering}

We use three approaches to filtering our simulated skymaps: the straightforward diagonal filtering of 
Eq.~\eqref{eqn:diagfilt}, on a masked map, diagonal filtering on the map with an apodized mask, and 
$C^{-1}$ filtering. 
The apodization and $C^{-1}$ procedures are described in more detail below.

\subsubsection{Apodization} 

\begin{figure}
\bc
\includegraphics[width=80mm,clip]{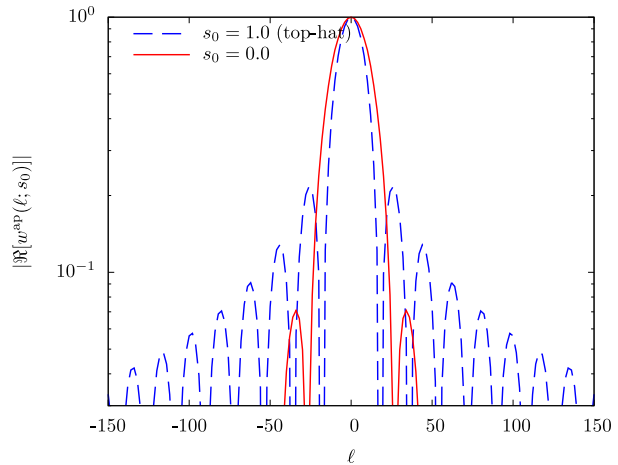}
\caption{
Real part of the Fourier counterpart of the apodizing function given in Eq.~(\ref{apo}) with $s_0=0.0$ 
(red solid line) compared with that of the top-hat function ($s_0=1.0$; blue dashed line). 
We choose $a=10$ deg. 
}
\label{Apo}
\ec
\end{figure}

One approach to reduce mode coupling from sky cuts, often used in power spectrum estimation 
(e.g., \citealt{Das:2007eu}), is apodization; to smooth the mask somewhat so that its Fourier 
counterpart more closely resembles a delta function. 
To apodize the survey boundary, for example, we can use a window function given by 
(as in Eq.~\ref{mask})
\al{
	W^{\rm s}(x,y;s_0)=w^{\rm s}(x;s_0)w^{\rm s}(y;s_0) 
	\,. 
}
We will use a sine apodization function given by
\al{
	w^{\rm s}(s;s_0) = \frac{1}{w_1} \times 
	\begin{cases} 1 & |s|<as_0 \\ \sin\left(\dfrac{\pi}{2}\dfrac{1-|s|}{1-s_0}\right) & as_0\leq |s|<a \\ 0 & a\leq |s| \end{cases}
	\,. \label{apo}
}
The parameter, $s_0$, indicates the width of the region where the apodization is applied, and the 
prefactor, $w_1\equiv2a[s_0+2(1-s_0)/\pi]$, is used so that $\int_{-\infty}^{\infty}\,ds\,w^{\rm s}(s,s_0)=1$. 
In Fig.~\ref{Apo}, we show the Fourier counterpart of the above function: 
\al{
	w^{\rm s}(\ell;s_0) = \int_{-\infty}^{\infty} ds \, e^{i\ell s}\,w^{\rm s}(s;s_0)
	\,. 
}
The Fourier counterpart with $s_0=0.0$ has a high-contrast peak at $\ell=0$ relative to that of the 
top-hat function ($s_0=1.0$). 
This implies that the function $W^{\rm s}(\hatn;s_0)$ given by Eq.~(\ref{apo}) with $s_0=0.0$ 
would be a better choice to reduce mode coupling, compared to that with $s_0=1.0$. 
From here forward, the fiducial value of 
parameter, $s_0$ in all of our survey boundary apodization is set to $s_0=0.0$. 
In order to see the dependence of this parameter, we also show the case with an intermediate value, $s_0=0.5$. 

We now construct an apodization function for both the survey boundaries and detected point sources. 
Let us consider an observed map given on an $S=[-a:a]\times[-a:a]$ plane with $N\rom{m}$ detected 
point sources, each of which we would like to mask. 
For simplicity, we will use a square mask function, with a length of $r\rom{m}$ on each size. 
We apply an apodizing window function to the observed map, given as
\al{
	W(\hatn;s_0,t_0) = \frac{1}{W_1(s_0,t_0)} W^{\rm s}(\hatn;s_0) \prod_{i=1}^{N\rom{m}} (1-W_{(i)}^{\rm m}(\hatn;t_0)) 
	\,, \label{apofunc}
}
The function, $W^{\rm s}(\hatn;s_0)$, is used to apodize the edges of the survey region, while the 
functions $1-W_{(i)}^{\rm m}(\hatn;t_0)$, apodize the point sources.
The factor $W_1$ is given by 
\al{
	W_1(s_0,t_0) = \int d^2\hatn \, e^{i\hatn\cdot\bl} W^{\rm s}(\hatn;s_0)\prod_{i=1}^{N\rom{m}} (1-W_{(i)}^{\rm m}(\hatn;t_0)) 
	\,, 
}
and the functions, $W_{(i)}^{\rm m}(\hatn;t_0)$ $(i=1,2,\dots,N\rom{m})$, are defined as 
\al{
	W_{(i)}^{\rm m}(x,y;t_0) = w_{(i)}^{\rm m}(x-x_i;t_0)w_{(i)}^{\rm m}(y-y_i;t_0)
	\,, 
}
with
\al{
	w_{(i)}^{\rm m}(t;t_0) = \begin{cases} 1 & |t|< b \\ 
		\sin\left(\dfrac{\pi}{2}\dfrac{b(1+t_0)-|t|}{bt_0}\right) & b\leq |t|<b(1+t_0) \\ 
		0 & b(1+t_0) \leq |t| 
		\end{cases}
	\,. \label{apo_mask}
}
The positions $(x_i,y_i)$ denote the position of $i$-th source mask, and $b=r\rom{m}/2$. 
The parameter, $t_0$, indicates the size of the apodization region for each source mask, and, 
similar to the case of $W^{\rm s}(\hatn;s_0)$, the Fourier counterpart, 
$W_{(i)}^{\rm m}(\bl;t_0)=\int d^2\hatn\,e^{i\bl\cdot\hatn}\,W_{(i)}^{\rm m}(\hatn;t_0)$, 
has a sharp peak at $\bl\sim \bm{0}$ for large values of $t_0$. 
If both functions, $W^{\rm s}(\hatn;s_0)$ and $W_{(i)}^{\rm m}(\hatn;t_0)$, are sharply peaked 
at $\bl\sim\bm{0}$ in Fourier space, the Fourier transform of $W(\hatn;s_0,t_0)$ can be approximated 
as a delta function.

\subsubsection{$C^{-1}$ filtering} 

The minimum-variance filtering which emerges from likelihood-based derivations of lensing 
estimators is known as $C^{-1}$ filtering. 
For the data model of Eq.~\eqref{eqn:datamodel} the inverse-variance filtered multipoles, 
$\bar{\Theta}_{\bl}$, are obtained by solving
\al{
	\left[1+\bm{C}^{1/2}\bm{N}^{-1}\bm{C}^{1/2}\right] (\bm{C}^{1/2}\bar{\bm{\Theta}}) 
		= \bm{C}^{1/2}\bm{N}^{-1} \hat{\Theta}
	\,, \label{Cinv}
}
where $\bar{\bm{\Theta}}$ is a vector whose components are $\bar{\Theta}_{\bl}$, $\bm{C}$ is 
the covariance of the lensed or unlensed CMB anisotropies with 
\al{
	\{\bm{C}\}_{\bl_i,\bl_j} &= \delta_{\bl_i-\bl_j} C_{\ell_i}^{\Theta\Theta} 
	\,, 
}
and $\bm{N}=\ave{\bm{n}^{\dagger}\bm{n}}$ is the covariance matrix for the instrumental noise. 
The noise covariance matrix in Fourier space is obtained from that in real space as 
\al{
	\bm{N}^{-1} = \bm{Y}^{\dagger}\bm{\ol{N}}^{-1}\bm{Y} 
	\,, 
}
where the pointing matrix, $\bm{Y}$, is defined by 
\al{
	\{\bm{Y}\}_{\hatn_i,\bl_j} = \exp(i\hatn_i\cdot\bl_j)
	\,. 
}
The mask is incorporated by setting the noise level of masked pixels to infinity, and therefore the 
inverse of the noise covariance in real space $\bm{\ol{N}}^{-1}$ to zero for masked pixels. 
The inversion of the matrix on the left-hand side of Eq.~(\ref{Cinv}) can be numerically costly, 
but may be evaluated using conjugate descent with careful preconditioning \citep{Smith07}. 

For our $C^{-1}$ results, after applying the $C^{-1}$ filter we additionally apply an apodizing 
function to account for the survey boundary, given by Eq.~(\ref{apo}).

\begin{figure*}
\bc
\includegraphics[width=80mm,clip]{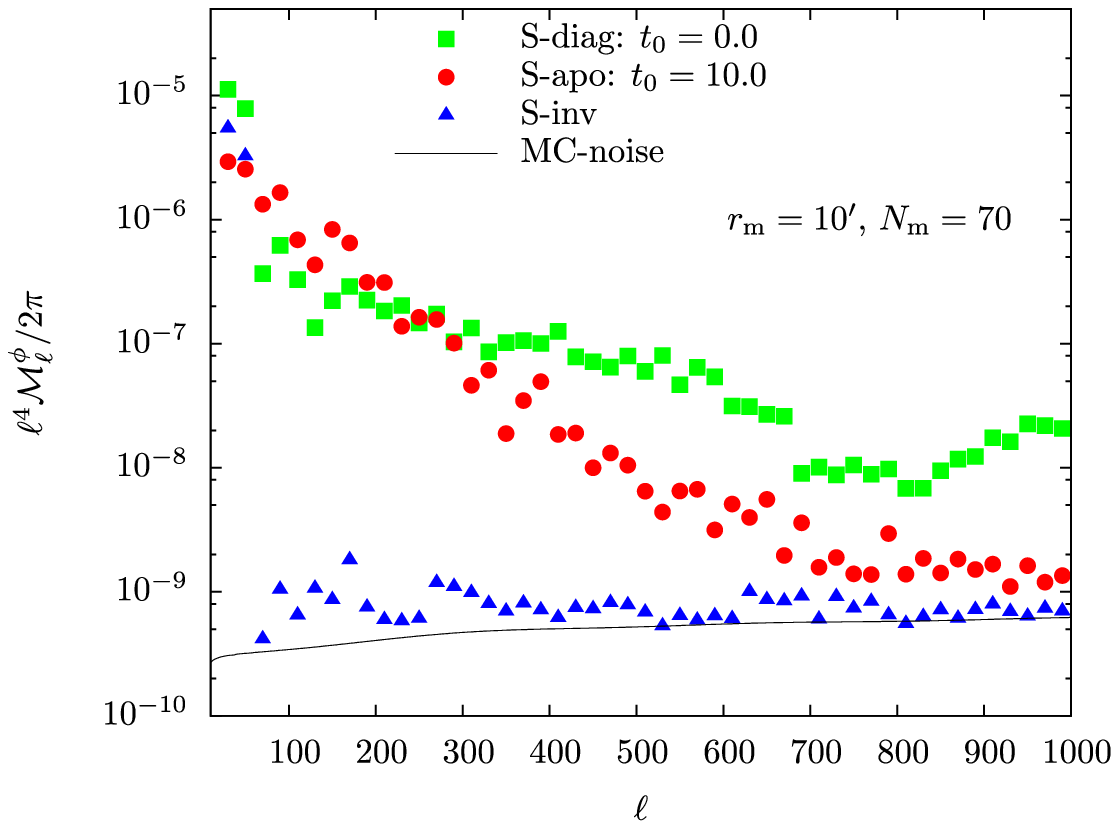}
\hspace{2em}
\includegraphics[width=80mm,clip]{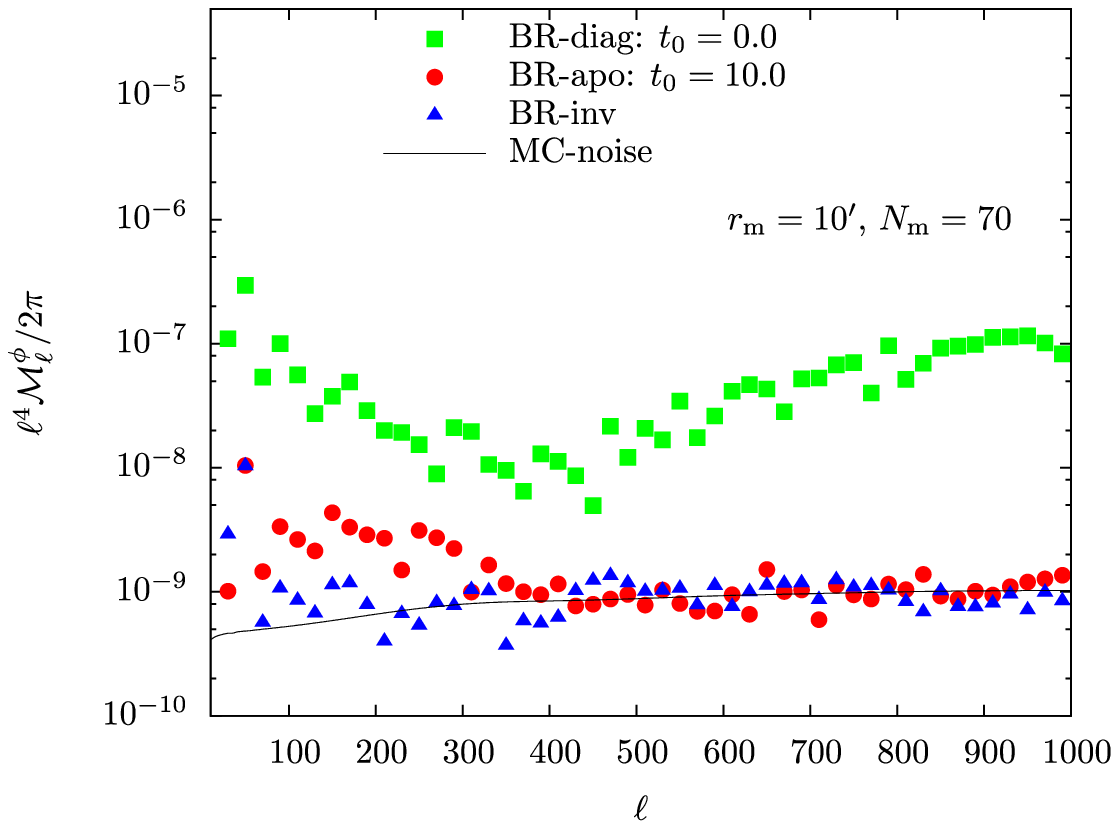}
\caption{
The power spectrum of the mask mean-field, using $100$ Gaussian unlensed map realizations with the 
standard quadratic estimator (left) and the bias-reduced estimator (right). 
In each panel, we show the case with diagonal filtering (S-diag, BR-diag), apodization (S-apo, BR-apo), 
and C-inverse filtered map (S-inv, BR-inv). 
The black line shows the Monte-Carlo noise, $N^{\grad,(0)}_{\ell}/100$.  
The multipoles are used up to $\ell\rom{max}=3000$. 
The number and size of masks, $N\rom{m}$ and $r\rom{m}$, are fixed with $70$ and $10'$, respectively; 
the total fraction of masked area is $\sim 2$\%. 
}
\label{Mean-G}
\ec
\end{figure*}

\begin{figure*}
\bc
\includegraphics[width=80mm,clip]{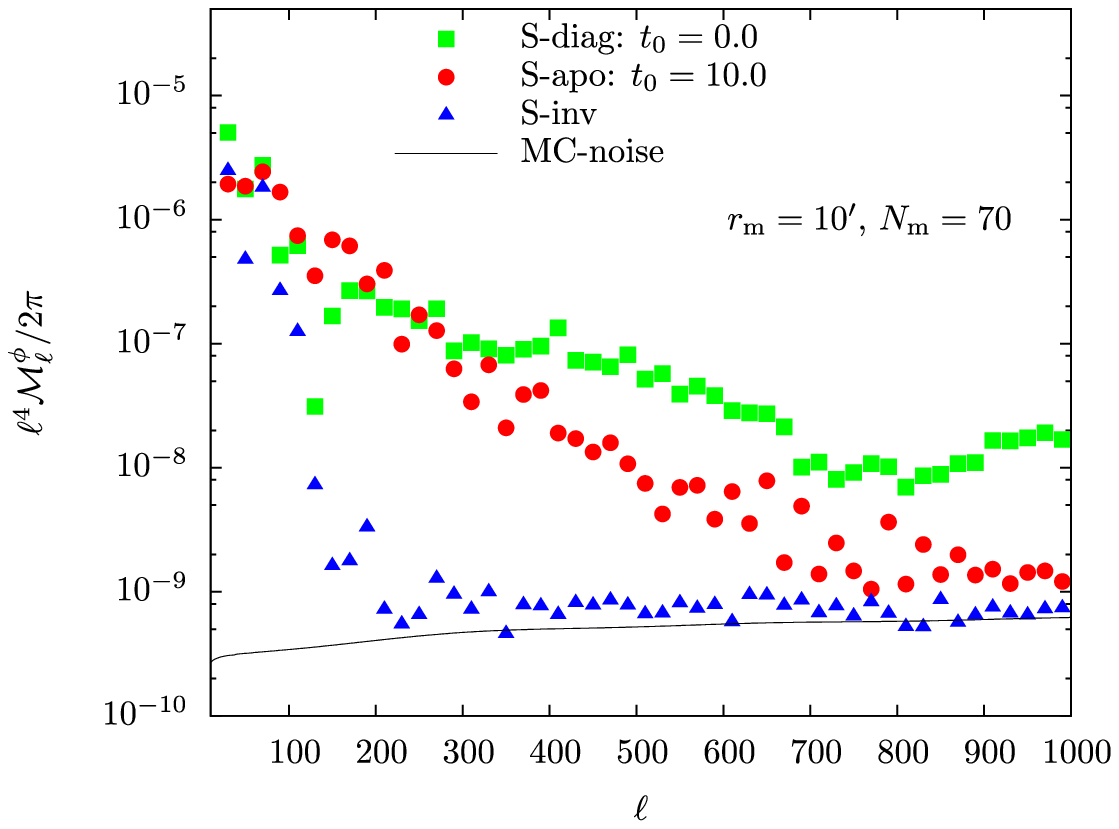}
\hspace{2em}
\includegraphics[width=80mm,clip]{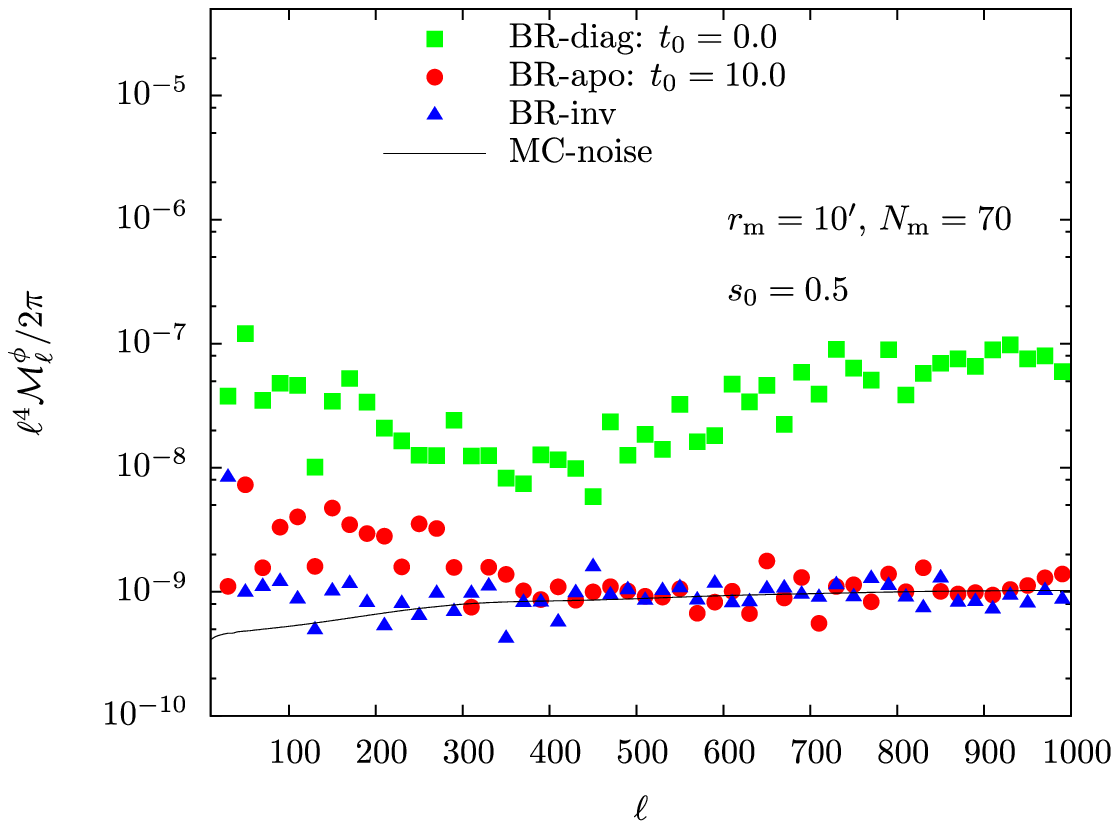}
\caption{
Same as Fig.~\ref{Mean-G} but for $s_0=0.5$. 
}
\label{Mean-G-1}
\ec
\end{figure*}

\begin{figure*}
\bc
\includegraphics[width=80mm,clip]{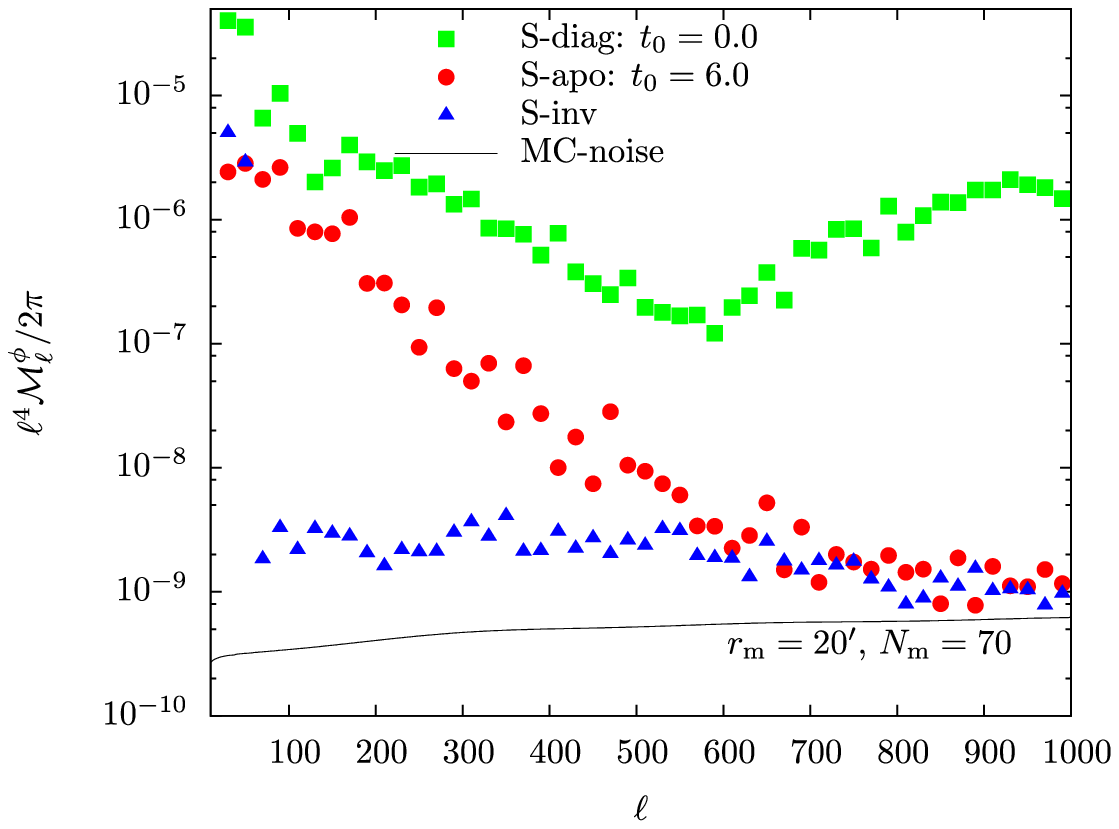}
\hspace{2em}
\includegraphics[width=80mm,clip]{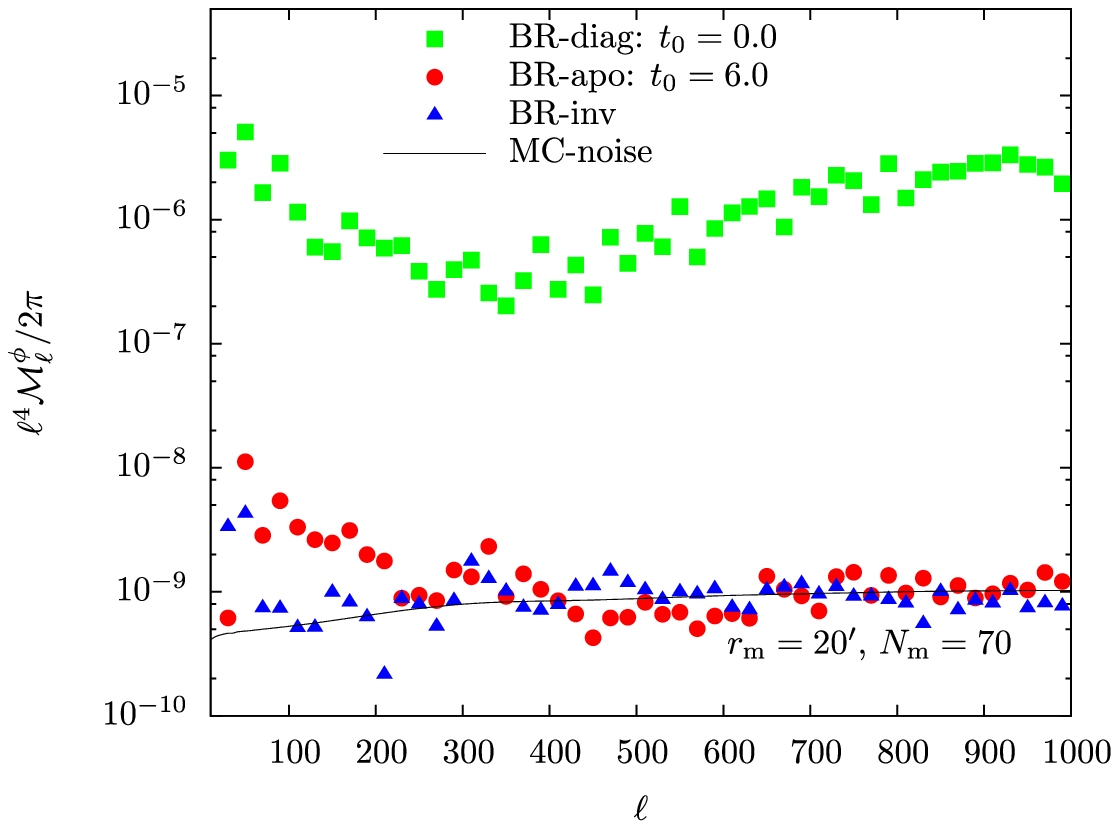}
\caption{
Same as Fig.\ref{Mean-G}, but for $r\rom{m}=20'$; 
the total fraction of masked area is $\sim 8$\%. 
}
\label{Mean-G-2}
\ec
\end{figure*}

\subsection{Mean-field power spectrum} 

\begin{figure*}
\bc
\includegraphics[width=80mm,clip]{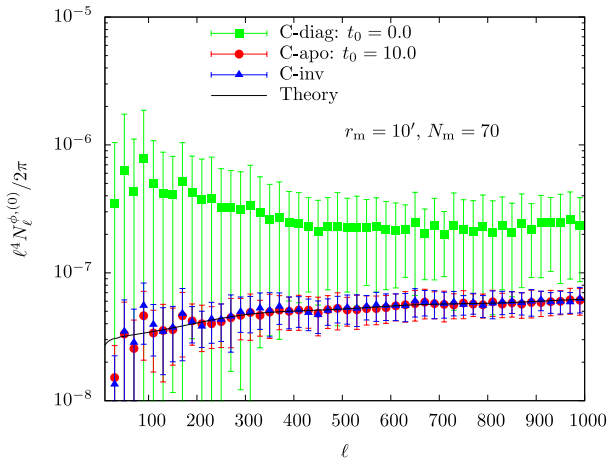}
\hspace{2em}
\includegraphics[width=80mm,clip]{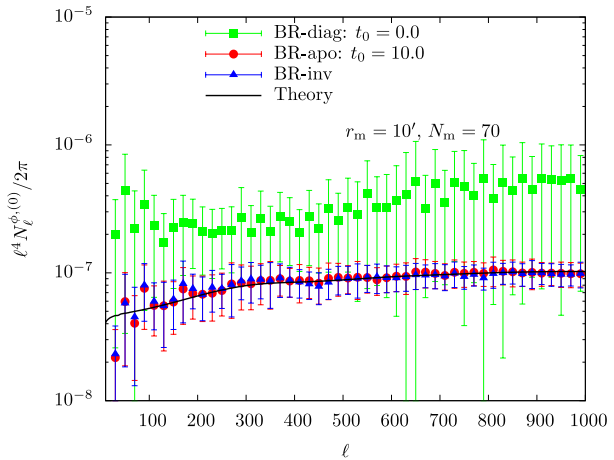} 
\caption{
The angular power spectrum of lensing estimator with Gaussian simulation, using the standard quadratic 
estimator with a perfect subtraction of the mean-field bias (Eq.~\ref{est}, left panel) and the bias-reduced estimator 
(Eq.~\ref{BR-est}, right panel). 
The mean and error bars of angular power spectrum are computed by $100$ realizations of the simulation. 
}
\label{GN-mr8}
\ec
\end{figure*}

We now proceed to our numerical results.
We start by looking at the power spectrum of the mean-field for both the standard and bias-reduced estimators. 

In Fig.~\ref{Mean-G}, we plot the power spectrum
\al{
	\mcM_{\ell}\equiv \frac{1}{W_2}\int\frac{d\varphi_{\bl}}{2\pi}
		\bigg|\frac{1}{100}\sum_{i=1}^{100}\hat{x}^{g,i}_{\bl} \bigg|^2
  \,, 
  \label{Mean-est}
}
where $\hat{x}^{g,i}_{\bl}$ is reconstructed from $i$-th realization of an unlensed Gaussian map 
without mean-field subtraction, and $W_2$ is derived (in analogy to Appendix~\ref{sec3.2.3}) as 
\al{
	W_2 &\equiv \int d^2\hatn \, W^2(\hatn; s_0, t_0) 
	\,. \label{W2}
}
We construct source masks with $N\rom{m}=70$ and $r\rom{m}=10'$. 
To show the usefulness of bias-reduced estimator, we compute 
\bi
\item the standard quadratic estimator (Eq.~\ref{est}), or 
\item the bias-reduced estimator (Eq.~\ref{BR-est}), 
\ei 
where the filtering is taken to be
\bi 
\item the diagonal filter with no apodization of source holes ($t_0=0$, denoted as S-diag) or
\item the diagonal filter with apodized source holes (non-zero $t_0$, denoted as S-apo), or 
\item the inverse-variance filtered map (denoted as S-inv). 
\ei
As noted above, in all cases we use $s_0=0.0$.
The results for the bias-reduced estimator are prefixed by ``BR''.

It is clear that the mean field bias from the standard quadratic estimator is large particularly on 
large scales, $\ell\lsim 500$. 
For the standard quadratic estimator, the mean-field on large scales still has a large amplitude even 
with source apodization or $C^{-1}$ filtering. 
When we use the inverse-variance filtering, the mean-field contribution from source holes is suppressed 
significantly. 
This is because the $\bm{\alpha}_{\bl}$ term in the estimator (c.f. Eq.~\ref{eqn:realspace}) in the case 
of $C^{-1}$ filtering corresponds to Wiener filtering, which is able to reconstruct the component of the 
masked signal which is due to modes larger than the holes themselves. 
Most of the power in the CMB gradient is due to scales greater than $r\rom{m}=10'$, and so this inpainting 
aspect of the $C^{-1}$ filter significantly reduces the generation of spurious gradients near the source 
boundary. 
Even for $C^{-1}$ filtering, however, there exists a mean-field on large scales $\ell\lsim 100$ due to 
the survey boundary in the standard estimator.
On the other hand, even without source apodization, the bias-reduced estimator suppresses the mean-field 
significantly. 
If we use the source apodization or $C^{-1}$ filtering, the mean-field is suppressed significantly, and 
the amplitude close to the Monte-Carlo noise level. 

Fig.~\ref{Mean-G-1} shows the case with $s_0=0.5$. 
The mean-field bias increases compared to the case with $s_0=0.0$, because of the residuals of 
survey boundary effect. 
This implies that, under the apodization function given in Eq.~(\ref{apo}), 
the mean-field bias from survey boundary effect would be minimized if we choose $s_0=0.0$. 

In Fig.~\ref{Mean-G-2}, we show the case with $r\rom{m}=20'$. 
Even in this case, for standard quadratic estimator, either the source apodization or $C^{-1}$ filtering 
suppresses the mean field significantly compared to the case without these two filtering methods, 
but, similar to $r\rom{m}=10'$, there are still large mean field at large scales. 
For the bias-reduced estimator, the mean field is suppressed down to the Monte-Carlo noise floor. 


\begin{figure}
\bc
\includegraphics[width=80mm,clip]{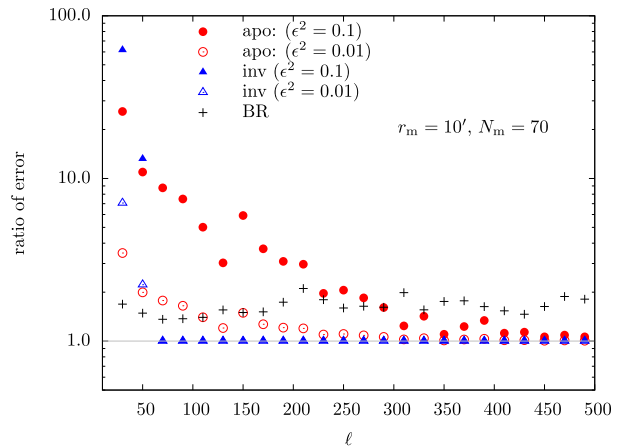}
\caption{
The error of estimator in the presence of uncertainties in the subtraction of mean-field bias. 
The points show the error in the case with apodization (red) or $C^{-1}$ filtering, 
assuming $\epsilon=0.1$ or $\epsilon=0.01$. 
For comparison, we also show the error of bias-reduced estimator. 
All errors are normalized by the case in the absence of uncertainties in the 
mean-field subtraction. 
}
\label{bias}
\ec
\end{figure}

\begin{figure}
\bc
\includegraphics[width=80mm,clip]{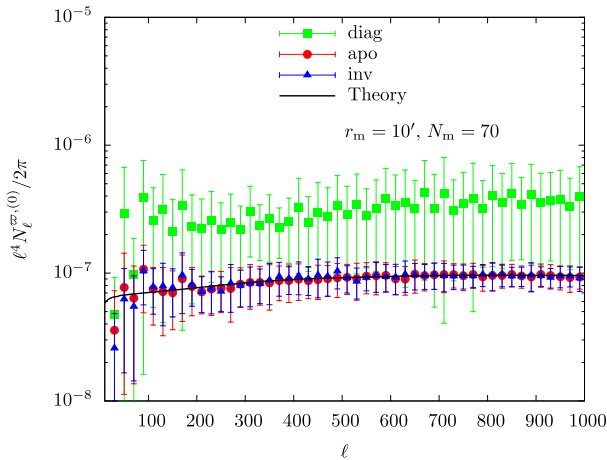}
\caption{
The angular power spectrum of the bias-reduced estimator for pseudo-scalar lensing potential, 
in the case with $N\rom{m}=70$ and $r\rom{m}=10'$. 
}
\label{GN-curl}
\ec
\end{figure}

\begin{figure*}
\bc
\includegraphics[width=80mm,clip]{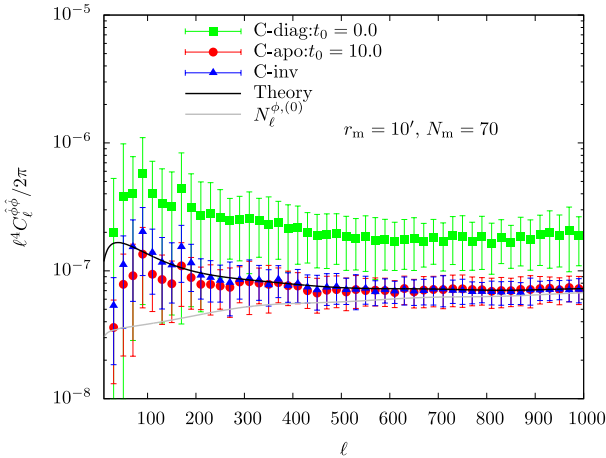}
\hspace{2em}
\includegraphics[width=80mm,clip]{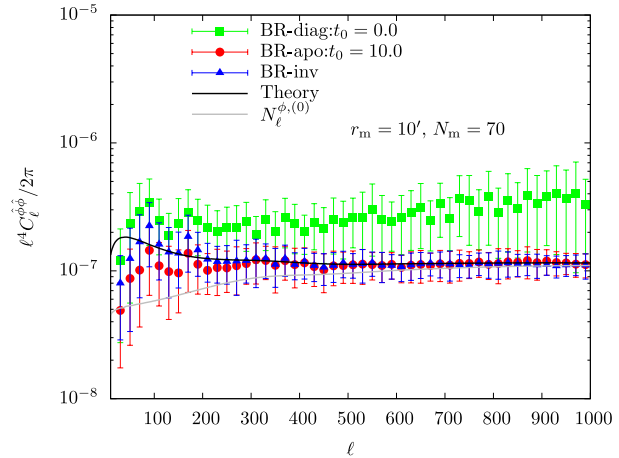} 
\caption{
Same as Fig.~\ref{GN-mr8}, but using simulated lensed temperature maps. 
The gray lines show the theoretical reconstruction noise bias, $N_{\ell}^{\grad,(0)}$. 
}
\label{LG-mr8}
\ec
\end{figure*}

\subsection{Power spectrum of lensing estimator} 

Next, we show the power spectrum of the lensing estimator computed from unlensed Gaussian 
simulations, i.e., the reconstruction noise bias. 
The reconstruction noise bias of the $i$-th realization map is computed as 
\al{
	\hat{N}_{\ell}^{xx,i} = \frac{1}{W_4}\int\frac{d\varphi_{\bl}}{2\pi}|\hat{x}^{g,i}_{\bl}|^2 
	\,, \label{Cl-est}
} 
and then the mean of $100$ realizations is compared with the full-sky, diagonal filtering expectation.
Note that, for the reduced bias estimator, the noise bias is modified, i.e., the standard reconstruction 
noise bias divided by \mbox{$(1-R_{\ell}^{\grad,M}R_{\ell}^{M,\grad})$}. 

In Fig.~\ref{GN-mr8}, we show the reconstruction noise bias, for the same cases as shown in 
Fig.~\ref{Mean-G}. 
We note that for both apodization and $C^{-1}$ filtering, the reconstruction noise bias agrees
well with the analytical approximation using either the standard estimator (after a perfect mean-field subtraction)
or the bias-reduced estimator. 

Now we turn to discuss the error of the estimators in the presence of incomplete mean-field subtraction 
in the conventional approach. 
The advantage of the bias-reduced approach is that it may be less strongly sensitive on how accurately 
we model the statistics of the underlying fluctuations. 
In the above analysis, although the error bars for the bias-reduced result are slightly larger than 
those for the standard result, the uncertainties in the mean-filed bias are completely ignored, 
and inclusion of other possible sources for mean-field bias degrades the accuracy in the standard approach. 
To see this, we consider a case with uncertainties in the mean-field bias generated by, e.g., 
calibration error as discussed in the previous section, and the resultant estimator is parametrized by 
a small parameter, $\epsilon$, as 
\beq 
	\hat{\grad}^{\rm C}_{\bl} \to \hat{\grad}^{\rm C}_{\bl} + \epsilon\ave{\hat{\grad}^{\rm S}_{\bl}} 
	\,. 
\eeq 
Fig.~\ref{bias} shows the errors taking into account this uncertainty normalized by the case for a perfect 
subtraction of mean-field bias, $1+\epsilon^2\mcM_{\ell}^{\grad}/N_{\ell}^{\grad,(0)}$, where 
the power spectrum of mean-field bias, $\mcM_{\ell}$, is computed for $N\rom{m}=70$ and $r\rom{m}=10'$. 
We also show the errors expected from the bias-reduced estimator divided by those from the standard approach 
without uncertainties in the mean-field subtraction. 
As clearly shown, the uncertainties for the standard approach are significant on large scales ($\ell\lsim100$), 
where the lensing signals are dominated, compared to that for the bias-reduced estimator. 


In Fig.~\ref{GN-curl}, we show the pseudo-scalar lensing potential. 
The reconstruction noise of the pseudo-scalar lensing potential is modified and the normalization is biased by 
the mode coupling due to the presence of sky cuts and masks. 
Note that there is no characteristic feature at large scales as there is for the scalar-lensing potential. 
This is because the estimator of the pseudo-scalar lensing potential is not significantly biased by the masking mean-field, 
as discussed in Sec.~\ref{sec.3}. 
Similar to the case with bias-reduced estimator of the scalar-lensing potential, using source apodization, 
the reconstruction noise agrees well with the analytical prediction for small source holes.

Finally, in Fig.~\ref{LG-mr8}, to see how well the bias can be reduced even in the presence of lensing field, 
as well as how well the estimator normalization is described by the full-sky equation, we show the angular 
power spectrum of lensing estimator, computed from Eq.~(\ref{Cl-est}) but using the estimator reconstructed 
from $i$-th realization of simulated lensed map. 
The theoretical prediction is the sum of the reconstruction noise, $N_{\ell}^{x,(0)} = A_{\ell}^{x}$, and 
the power spectrum of scalar-lensing potential, $C_{\ell}^{\grad\grad}$. 
The results are similar to that in the case with the unlensed Gaussian simulation.

\section{Summary} 
\label{sec.5} 

We have discussed methods for removing the ``mean-field'' and ``reconstruction noise'' biases 
which must be accounted for in CMB lens reconstruction.
Our approach focuses on estimating these biases directly from the data itself, reducing our sensitivity 
to the sky model which is otherwise needed to determine them.
We performed numerical tests of the mean-field reduction approach for several different choices 
of filtering, finding it particularly useful for the reduction of the large-scale component of the mean-field. 

In our analysis, we have focused on the temperature fluctuations, but upcoming and next-generation 
CMB experiments will also provide information on the polarization. 
The polarization anisotropies are a more sensitive probe of lensing effects and thus the lensing 
reconstruction from realistic polarization maps is also worth investigating. 
The method investigated in this paper may be also applicable to the polarization maps, and the 
usefulness of our method to the lensing reconstruction from polarization maps will be explored in 
our future work (Namikawa et al in prep.). 

Our simulation results of the lensed temperature anisotropies are available as numeric
tables upon request (contact takahasi@cc.hirosaki-u.ac.jp).


\section*{Acknowledgments}
We thank Aur\'{e}lien Benoit-L\'{e}vy, Karim Benabed, Ryo Nagata, and 
Eiichiro Komatsu for useful comments and discussions, 
and greatly appreciate Takashi Hamana and Takahiro Nishimichi for kindly 
providing the ray-tracing simulation code and the 2LPT code. 
TN is also grateful to Atsushi Taruya for several comments and discussions. 
This work was supported in part by Grant-in-Aid for Scientific Research 
on Priority Areas No. 467 
``Probing the Dark Energy through an Extremely Wide and Deep Survey with Subaru Telescope'', 
JSPS Core-to-Core Program 
``International Research Network for Dark Energy'', 
Hirosaki University Grant for Exploratory Research by Young Scientists, 
by the Grand-in-Aid for the Global COE Program 
``Quest for Fundamental Principles in the Universe: from Particles
to the Solar System and the Cosmos'' from the Ministry of Education,
Culture, Sports, Science and Technology (MEXT) of Japan,
by the MEXT Grant-in-Aid for Scientific Research on Innovative Areas
(No. 21111006), by the FIRST program 
``Subaru Measurements of Images and Redshifts (SuMIRe)'', 
World Premier International Research Center Initiative (WPI Initiative) 
from MEXT of Japan. 
Numerical computations were carried out on SR16000 at YITP in Kyoto
University and Cray XT4 at Center for Computational Astrophysics,
CfCA, of National Astronomical Observatory of Japan.

\appendix
\section{
Numerical simulation of lensed CMB maps 
} 
\label{appA} 

In this section, we describe our method to generate lensed CMB maps, 
based on the ray-tracing of large-scale structure simulations 
(e.g., \citealt{Sato:2009ct,Takahashi:2011qd}). 

\begin{figure}
\bc
\vspace*{1.0cm}
\includegraphics[width=70mm]{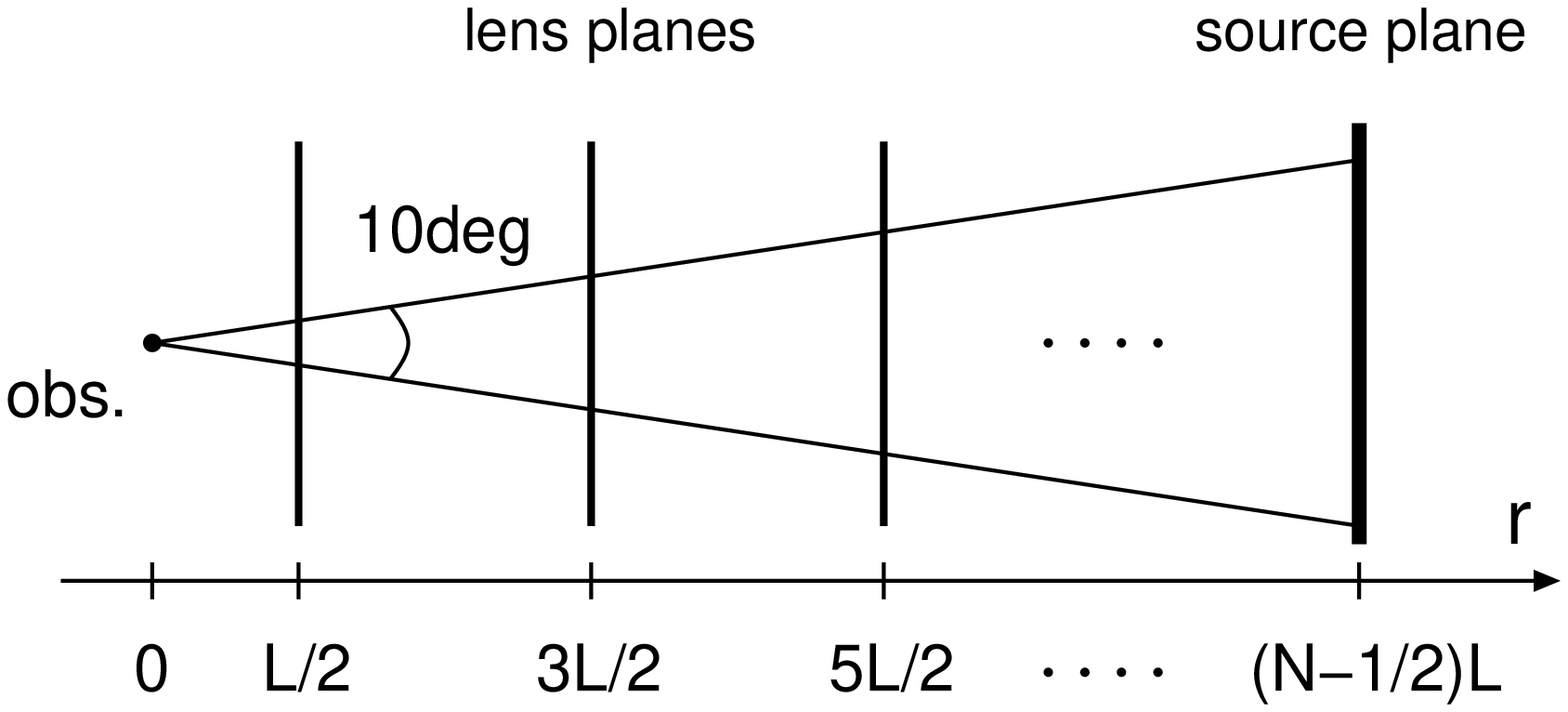}
\caption{
A schematic picture of our ray-tracing simulation. 
The horizontal axis is the comoving distance $r$ from the observer. 
The vertical thick lines denote the positions of the lens planes and 
the source plane, which are located at $r=(i-1/2) \times L$ with 
$i=1,2,\cdots,N$. 
Here, we set $N=20$. 
We have multiple lens planes but use the flat-sky approximation. 
Light rays are emitted from the observer and are deflected at the lens 
planes before reaching the source plane. 
The field of view is $10 \times 10$ deg$^2$.
}
\label{fig_lens_planes}
\vspace*{0.5cm}
\ec
\end{figure}

Fig.~\ref{fig_lens_planes} shows a schematic picture of our ray-tracing 
simulation. 
We have multiple lens planes but use a flat-sky approximation. 
The horizontal axis is the comoving distance $r$ from the observer. 
The thick vertical lines are the lens and source planes, which are placed 
at equal-distance intervals of $L$, and located at $r=(i-1/2) \times L$ 
with an integer $i=1,2,\cdots,N$. 
Here, $i=N$ corresponds to the source plane. 
We place the source plane at the last scattering surface $(z_s=1090)$. 
The distance to the last scattering surface is 
$r_{\rm LSS}=r(z_s=1090)=9900h^{-1}$Mpc in our fiducial cosmological model. 
We place $19$ lens planes up to the last scattering surface, i.e., 
we set $N=20$. 
We determine the interval $L$ from the distance to the last scattering 
surface divided by the number of lens planes, 
$L=r_{\rm LSS}/(N-1/2)=r_{\rm LSS}/(20-1/2)=507.6h^{-1}$Mpc. 
Light rays are emitted from the observer and are deflected at each lens 
plane before reaching the source plane. 
In our simulation, the field of view is $10\times 10$ deg$^2$. 
We impose a periodic boundary condition on the lens planes. 

\subsection{$N$-body Simulations}

In order to obtain the particle distribution and the gravitational 
potential on the lens planes, we run $N$-body simulations in a 
cubic box, and then project the particle positions into two dimensions perpendicular to the line-of-sight. 
We use the numerical simulation code Gadget2 \citep{Springel:2000yr,Springel:2005mi}. 
We generate the initial conditions based on the second-order Lagrangian 
perturbation theory (2LPT; \citealt{Crocce:2006ve,Nishimichi:2009}) with 
the initial linear power spectrum calculated by CAMB. 
We employ $1024^3$ dark matter particles in the simulation box of 
$L=507.6$ $h^{-1}$Mpc on a side. 
The initial redshift is $z_{\rm init}=70$, and we dump the outputs 
(the particle positions) at the redshifts corresponding to the positions 
of the lens planes $r=L \times (i-1/2)$, shown in Fig.~\ref{fig_lens_planes}.
The softening length is fixed to be $5\%$ of the mean particle separations, 
which correspond to $25$ $h^{-1}$kpc. 
We prepare five independent realizations to reduce the sample variance. 

\subsection{Ray-tracing Simulations}
We briefly explain the procedure to trace light rays through $N$-body 
data and obtain the maps of the lensing fields on the source plane 
(see also, e.g., \citealt{Sato:2009ct,Takahashi:2011qd}). 
We use the code RAYTRIX \citep{Hamana:2001vz} which follows the standard 
multiple lens plane algorithm. 
In the standard multiple lens plane algorithm, the distance between 
observer and source galaxies is divided into several intervals. 
In our case, as shown in Fig.~\ref{fig_lens_planes}, we adopt a fixed 
interval whose value is the same as simulation box $L$ on a side. 
Particle positions are projected onto two dimensional lens planes 
($xy$, $yz$ ,$zx$ planes) every $L$. 
Using the Triangular-Shaped Cloud method \citep{Hockney:1988}, we assign 
the particles onto $N_g^2$ grids in lens planes, then compute the 
projected density contrast at each plane. 
We use $N_g^2=2048^2$ throughout this paper. 
The two-dimensional gravitational potential is solved via the Poisson 
equation using a Fast Fourier Transform. 
Finally, two dimensional sky maps of the convergence, shear, and 
angular positions of light rays are obtained by solving the evolution 
equation of Jacobian matrix along the light-ray path which is obtained 
by solving the multiple lens equation. 
At high redshifts ($z>12$), the density fluctuations are very small and 
give only $<10\%$ contribution to the angular power spectrum of the lensing 
potential at the multipole $\ell=100-1000$ (see \citealt{Lewis:2006fu}).
Hence, at high redshifts ($z>12$) we do not include the density 
fluctuations in our calculation. 
We solve the multiple lens equation up to $z=12$ and further redshifts 
the light rays are assumed to propagate in straight lines. 

We prepare $100$ realizations by randomly choosing the projecting 
direction and shifting the two dimensional positions. 
In each realization, we emit $1024^2$ light-rays in the field-of-view 
$10 \times 10$ deg$^2$. 
Then, the angular resolution is $10$deg$/1024 \simeq 0.6'$. 

\begin{figure}
\bc
\vspace*{1.0cm}
\includegraphics[width=70mm]{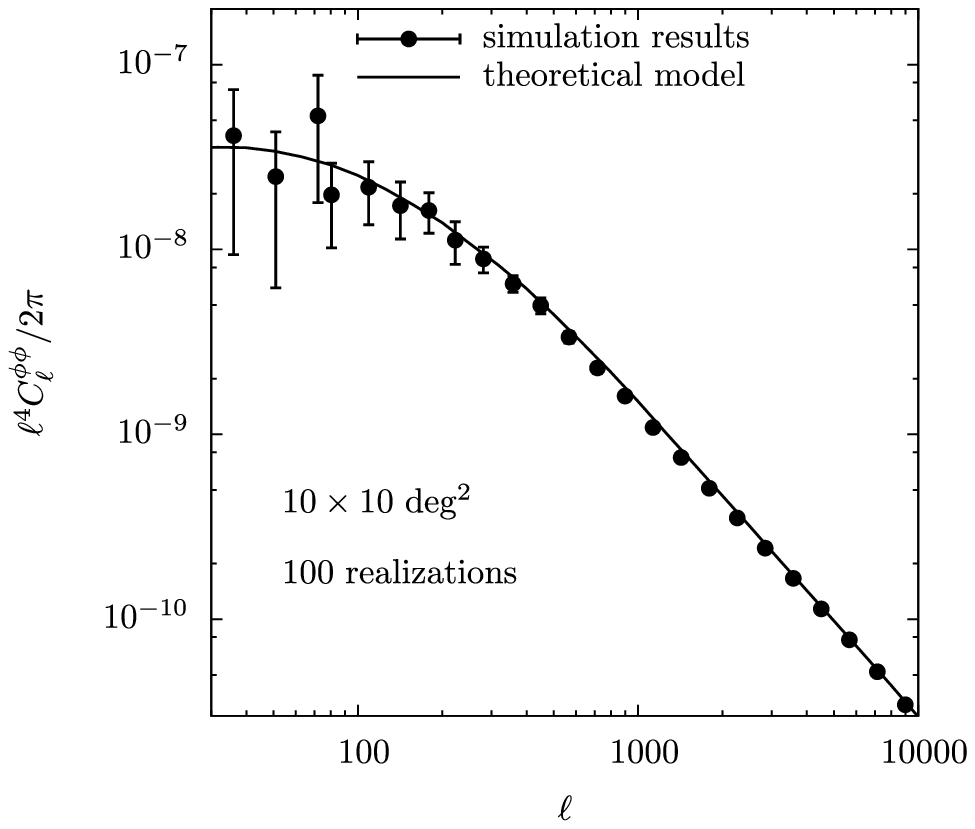}
\caption{
The power spectrum of the scalar lensing potential 
at the last scattering surface ($z=1090$). 
The dots with error bars are our ray-tracing simulation results 
calculated from the $100$ retaliations of $10\times 10$ deg$^2$ 
convergence maps. The black curve is the theoretical prediction. 
}
\label{fig_conv_cl}
\vspace*{0.5cm}
\ec
\end{figure}

In order to check the accuracy of our ray-tracing simulation, we calculate 
the power spectrum of scalar lensing potential and compare it with the 
theoretical model. 
Fig.~\ref{fig_conv_cl} shows the power spectrum, 
$\ell^4C_{\ell}^{\grad \grad}/2\pi$, as a function of multipole $\ell$ 
at the last scattering surface ($z=1090$). 
The dots with error bars are the simulation results calculated from the 
$100$ realizations of $10 \times 10$ deg$^2$ convergence maps. 
The black curve is the theoretical prediction in which the power spectrum 
of scalar lensing potential is given by the projected (three-dimensional) 
matter power spectrum weighted with the radial lensing kernel along the 
line-of-sight (e.g., \citealt{Bartelmann:1999yn}). 
Here, we use the halo-fit model 
\citep{Smith:2002dz} to calculate the non-linear power spectrum. 
As clearly seen in the figure, our simulation results agree with the 
theoretical model very well.

\subsection{Lensed CMB Temperature Map}

\begin{figure}
\bc
\vspace*{1.0cm}
\includegraphics[width=70mm]{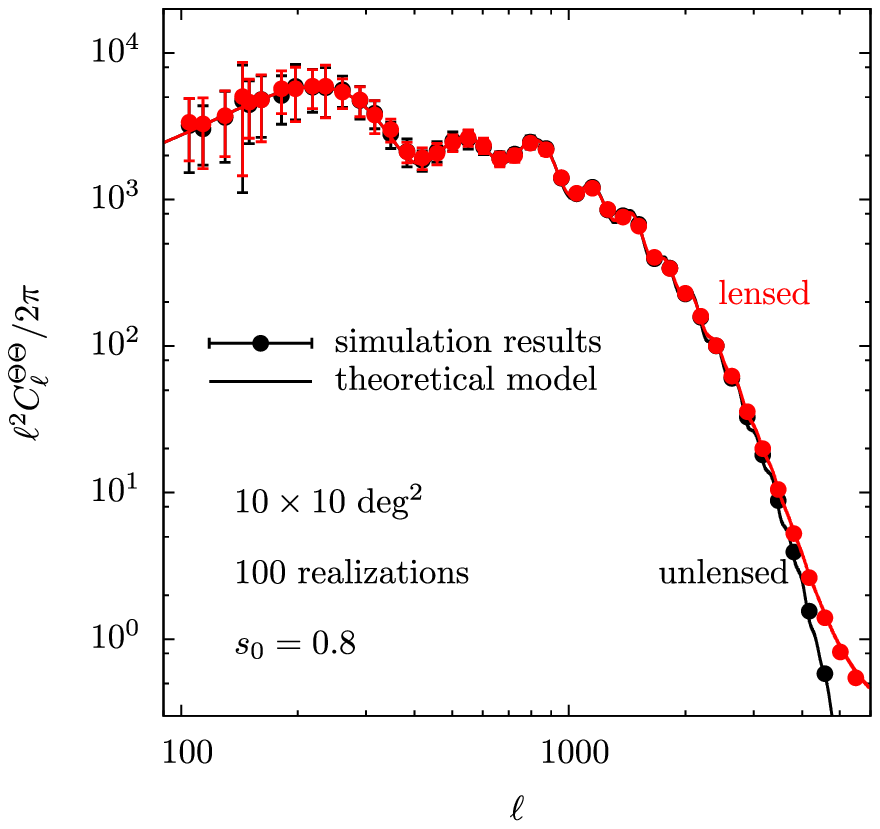}
\caption{
The lensed CMB temperature power spectrum. 
The red symbols are the lensed power spectrum, while the black symbols 
are unlensed one. The dots with error bars are our simulation results 
calculated from the $100$ realizations of $10 \times 10$ deg$^2$ lensed maps. 
Here, we use $s_0=0.8$ in the apodization. 
The solid curves are the results from CAMB. 
}
\label{fig_cmb_cl}
\vspace*{0.5cm}
\ec
\end{figure}

In this subsection, we introduce our procedure for making  lensed CMB 
temperature maps. 
We prepare these maps as follows:
\begin{enumerate}
\item We obtain the power spectrum of the unlensed CMB temperature
    fluctuations using CAMB.
\item We make an unlensed temperature map on a square $\sqrt{4 \pi}$
    radian ($\simeq 203$ deg) on a side assuming Gaussian fluctuations 
    based on the unlensed power spectrum. 
    The angular resolution of the temperature fluctuations is
    set to be $10$ deg$/1024 \simeq 0.6'$.
    We prepare $100$ such unlensed maps.
\item Finally, we calculate the deflection angle ${\bm{d}}(\hat{\bm{n}})$ 
    at the angular position $\hat{\bm{n}}$ using the ray-tracing simulation
    for the $1024^2$ light rays.
    Then, we obtain the lensed temperature map by shifting
    the positions $\hat{\bm{n}} \rightarrow \hat{\bm{n}} +
    {\bm{d}}(\hat{\bm{n}})$ on the unlensed map, according to Eq.~(\ref{remap}). 
    We have $100$ lensed CMB temperature maps of $10 \times 10$ deg$^2$ 
    with $1024^2$ grids. 
\end{enumerate}
We calculate the power spectrum from the $100$ lensed CMB temperature maps, and 
the result is shown in Fig.~\ref{fig_cmb_cl}. 
The figure shows the angular power spectrum of lensed temperature fluctuations 
as a function of multipole $\ell$. 
The dots with error bars are the mean 
values and the dispersions calculated from the $100$ realizations. 
We use $s_0=0.8$ in the apodization. 
The red symbols are the lensed power spectrum, while the black symbols are unlensed. 
The solid curves are the theoretical prediction calculated by CAMB. 
Our simulation results agree with the theoretical predication very well.

\section{Apodized Normalization} \label{sec3.2.3} 

Apodization/masking leads to a change in the angular power spectrum of the lensing estimator.
Here we derive the appropriate correction to the normalization of the estimator power spectrum.

Substituting Eq.~(\ref{fourier-W}) into Eq.~(\ref{est}), the estimator in the presence of an general 
window function is given as 
\al{
	\hat{x}_{\bl} &= \Int{\bL}{F}^x_{\bl,\bL}\Int{\bl_1\p} \Int{\bl_2\p} W_{\bL-\bl_1\p} W_{\bl-\bL-\bl_2\p} 
	\hat{\Theta}_{\bl_1\p}\hat{\Theta}_{\bl_2\p}
	\,, \label{W-x}
}
where we define ${F}^x_{\bl,\bL}={A}_{\ell}^{x}{f}^x_{\bl,\bL}/2\hat{\mcC}_L^{\Theta\Theta}\hat{\mcC}_{|\bl-\bL|}^{\Theta\Theta}$. 
From the above equation, the angular power spectrum of the estimator becomes 
\begin{multline}
	\ave{|\hat{x}_{\bl}|^2} = \Int{\bL}\Int{\bL\p}{F}^x_{\bl,\bL}{F}^x_{\bl,\bL\p}
	\\ \times \Int{\bl_1}\Int{\bl_2}\Int{\bl_3}\Int{\bl_4}
	\\ \times W_{\bL-\bl_1}W_{\bl-\bL-\bl_2}W_{\bL-\bl_3}W_{\bl-\bL-\bl_4}
	\\ \times
	\ave{\hat{\Theta}_{\bl_1}\hat{\Theta}_{\bl_2}\hat{\Theta}_{-\bl_3}\hat{\Theta}_{-\bl_4}}
	\,, 
\end{multline}
where we use $\hat{\Theta}^*_{\bL}=\hat{\Theta}_{-\bL}$. 
Note that the statistical anisotropy of lensed fields restrict the integral of Fourier modes, $\bl_1$, $\bl_2$, $\bl_3$ and $\bl_4$, to 
the case with $\bl_1+\bl_2-\bl_3-\bl_4=\bm{0}$, the delta function, $\delta_{\bl_1+\bl_2-\bl_3-\bl_4}$, is multiplied in the integrand of 
the above equation. 
If the window functions behave as the delta function, the above equation reduces to 
\begin{multline}
	\ave{|\hat{x}_{\bl}|^2} \simeq \Int{\bL}\Int{\bL\p}{F}^x_{\bl,\bL}{F}^x_{\bl,\bL\p}
	\\ \times
	\ave{\hat{\Theta}_{\bL}\hat{\Theta}_{\bl-\bL}\hat{\Theta}_{-\bL}\hat{\Theta}_{-\bl+\bL}}W_4 
	\,. \label{W4}
\end{multline}
The quantity $W_4$ is defined by 
\begin{multline}
	W_4 \equiv \Int{\bl_1}\Int{\bl_2}\Int{\bl_3}\Int{\bl_4}
	\\ \times
	W_{\bL-\bl_1}W_{\bl-\bL-\bl_2}W_{\bL-\bl_3}W_{\bl-\bL-\bl_4}\delta_{\bl_1+\bl_2-\bl_3-\bl_4} 
\end{multline}
which reduces to
\begin{eqnarray}
	W_4 &=& \prod_{i=1}^4\Int{\bl_j\p}W_{\bl_j\p}\int d^2\hatn \, e^{i\hatn\cdot(\bl_1+\bl_2-\bl_3-\bl_4)} \nonumber \\
	&=& \int d^2\hatn \, W^4(\hatn) 
	\,. 
\end{eqnarray}
Eq.~(\ref{W4}) means that the power spectrum of reconstructed estimator from finite-size masked 
map is equal to that from ideal map multiplied by the quantity $W_4$.

\bibliographystyle{mn2e}
\bibliography{cite}

\end{document}